\documentclass[fleqn,usenatbib,useAMS]{mnras}

\usepackage{graphicx}	
\usepackage{amsmath}	
\usepackage{amssymb}	
\usepackage{multicol}        
\usepackage{bm}		
\usepackage{pdflscape}	
\usepackage{mathtools}
\usepackage{multirow}
\usepackage{xcolor}
\usepackage[utf8]{inputenc}
\usepackage[T1]{fontenc}
\usepackage{ae,aecompl}

\newcommand{\NV}{\mbox{N \textsc{v}}}
\newcommand{\SiIV}{\mbox{Si \textsc{iv}}}
\newcommand{\CIV}{\mbox{C \textsc{iv}}}
\newcommand{\CII}{\mbox{C \textsc{ii}}}
\newcommand{\MgII}{\mbox{Mg \textsc{ii}}}
\newcommand{\SII}{\mbox{S \textsc{ii}}}
\newcommand{\ZnII}{\mbox{Zn \textsc{ii}}}
\newcommand{\TiII}{\mbox{Ti \textsc{ii}}}
\newcommand{\FeII}{\mbox{Fe \textsc{ii}}}
\newcommand{\AlII}{\mbox{Al \textsc{ii}}}
\newcommand{\AlIII}{\mbox{Al \textsc{iii}}}
\newcommand{\SiII}{\mbox{Si \textsc{ii}}}
\newcommand{\OI}{\mbox{O \textsc{i}}}

\newcommand{\CoII}{\mbox{Co \textsc{ii}}}
\newcommand{\CrII}{\mbox{Cr \textsc{ii}}}
\newcommand{\CaII}{\mbox{Ca \textsc{ii}}}
\newcommand{\MnII}{\mbox{Mn \textsc{ii}}}
\newcommand{\NaI}{\mbox{Na \textsc{i}}}
\newcommand{\MgI}{\mbox{Mg \textsc{i}}}
\newcommand{\Lya}{Ly$\alpha$}
\newcommand{\numCIV}{479}
\newcommand{\numMgII}{360}
\newcommand{\numCII}{46}
\newcommand{\kms}{\,km\,s$^{-1}$}

\newcommand{\appropto}{\mathrel{\vcenter{
  \offinterlineskip\halign{\hfil$##$\cr
    \propto\cr\noalign{\kern2pt}\sim\cr\noalign{\kern-2pt}}}}}

\newenvironment{nscenter}
 {\parskip=0pt\par\nopagebreak\centering}
 {\par\noindent\ignorespacesafterend}

\title[XQR-30 Metal Absorber Catalog]{The XQR-30 Metal Absorber Catalog: 778 Absorption Systems Spanning 2 $\lesssim z \lesssim$ 6.5}

\author[Rebecca L. Davies et al.]{\href{https://orcid.org/0000-0002-3324-4824}{Rebecca L. Davies}$^{1,2}$\thanks{Contact e-mail: \href{mailto:rdavies@swin.edu.au}{rdavies@swin.edu.au}}, %
\href{https://orcid.org/0000-0002-5360-8103}{E. Ryan-Weber}$^{1,2}$, %
\href{https://orcid.org/0000-0003-3693-3091}{V. D'Odorico}$^{3,4,5}$, %
\href{https://orcid.org/0000-0001-8582-7012}{S. E. I. Bosman}$^6$, %
\href{https://orcid.org/0000-0001-5492-4522}{R. A. Meyer}$^6$, %
\newauthor \href{https://orcid.org/0000-0003-2344-263X}{G. D. Becker}$^7$, %
\href{https://orcid.org/0000-0002-6830-9093}{G. Cupani}$^3$, %
\href{https://orcid.org/0000-0002-4314-021X}{M. Bischetti}$^{3,8}$, %
\href{https://orcid.org/0000-0003-0875-9911}{A. M. Sebastian}$^{1,2}$, %
\href{0000-0003-2895-6218}{A.-C.~Eilers}$^9$\textdagger, \href{0000-0002-6822-2254}{E. P. Farina}$^{10}$, %
\newauthor \href{0000-0002-7633-431X}{F. Wang}$^{11}$\textdagger, \href{0000-0001-5287-4242}{J. Yang}$^{11}$\ddag, \href{0000-0003-3307-7525}{Y. Zhu}$^7$ \\
$^1$Centre for Astrophysics and Supercomputing, Swinburne University of Technology, Hawthorn, Victoria 3122, Australia \\
$^2$ARC Centre of Excellence for All Sky Astrophysics in 3 Dimensions (ASTRO 3D), Australia \\
$^3$INAF-Osservatorio Astronomico di Trieste, Via Tiepolo 11, I-34143 Trieste, Italy \\
$^4$IFPU-Institute for Fundamental Physics of the Universe, via Beirut 2, I-34151 Trieste, Italy \\
$^5$Scuola Normale Superiore, Piazza dei Cavalieri 7, I-56126 Pisa, Italy \\
$^6$Max-Planck-Institut f\"ur Astronomie, K\"onigstuhl 17, D-69117 Heidelberg, Germany \\
$^7$Department of Physics \& Astronomy, University of California, Riverside, CA 92521, USA \\
$^8$Dipartimento di Fisica, Sezione di Astronomia, Universit\'a di Trieste, via Tiepolo 11, 34143 Trieste, Italy \\
$^9$MIT Kavli Institute for Astrophysics and Space Research, 77 Massachusetts Ave., Cambridge, MA 02139, USA \\
$^{10}$Gemini Observatory, NSF’s NOIRLab, 670 N A’ohoku Place, Hilo, Hawai'i 96720, USA \\
$^{11}$Steward Observatory, University of Arizona, 933 N Cherry Ave., Tucson, AZ 85721, USA \\
\textdagger NASA Hubble Fellow\\
\ddag Strittmatter Fellow
}

\pubyear{2022}

\begin{document}
\label{firstpage}
\pagerange{\pageref{firstpage}--\pageref{lastpage}}
\maketitle

\begin{abstract}
Intervening metal absorption lines in the spectra of $z\gtrsim$~6 quasars are fundamental probes of the ionization state and chemical composition of circumgalactic and intergalactic gas near the end of the reionization epoch. Large absorber samples are required to robustly measure typical absorber properties and to refine models of the synthesis, transport, and ionization of metals in the early Universe. The ``Ultimate XSHOOTER legacy survey of quasars at $z\sim$~5.8-6.6'' \mbox{(XQR-30)} has obtained high signal-to-noise spectra of 30 luminous quasars, nearly quadrupling the existing sample of 12 high quality $z\sim$~6 quasar spectra. We use this unprecedented sample to construct a catalog of 778 systems showing absorption in one or more of \MgII\ (\numMgII\ systems), \FeII\ (184), \CII\ (\numCII), \CIV\ (\numCIV), \SiIV\ (127), and \NV\ (13) which span \mbox{2~$\lesssim z \lesssim$ 6.5}. This catalog significantly expands on existing samples of \mbox{$z \gtrsim$~5} absorbers, especially for \CIV\ and \SiIV\ which are important probes of the ionizing photon background at high redshift. The sample is 50\% (90\%) complete for rest-frame equivalent widths $W \gtrsim$~0.03\AA\ (0.09\AA). We publicly release the absorber catalog along with completeness statistics and a \textsc{Python} script to compute the absorption search path for different ions and redshift ranges. This dataset is a key legacy resource for studies of enriched gas from the era of galaxy assembly to cosmic noon, and paves the way for even higher redshift studies with the \textit{James Webb Space Telescope} and 30m-class telescopes.
\end{abstract}


\section{Introduction}
High redshift quasars are extremely valuable tools for studying the properties of the Universe during the later stages of the epoch of reionization ($z\sim$~6). Intervening metal absorption lines in quasar spectra encode the redshift, chemical content, and ionization state of gas in and around galaxies along the line of sight. The rest-frame equivalent widths of absorption lines do not decrease as the quasar light moves through space, providing a means to robustly detect and study faint galaxies in the distant Universe that are below the flux limit of current state-of-the-art emission line surveys. The chemical abundance ratios reflect the properties of the stars that synthesized the metals, offering key insights into the nature of early stars and galaxies \citep[e.g.][]{Cooke11, Becker12, Banados19, Welsh22}. The ionization state of the absorbing gas is sensitive to the shape and strength of the ionizing radiation field, providing independent constraints on the timing and sources of reionization \citep[e.g.][]{Finlator16, Garcia17b, Doughty18, Gotberg20}. The overall incidence of metal absorbers traces the average metal content of galaxies and the circumgalactic medium (CGM), and can be used to test models of the production and transport of metals in the early Universe (see \citealt{Becker15Review} and \citealt{Tumlinson17} for reviews). 

Over the last decade, the increasing availability of $z \gtrsim$~6 quasar spectra has propelled our understanding of the Universe during its infancy. Measurements of \Lya\ forest dark gaps \citep[e.g.][]{Songaila02, Furlanetto04, Gallerani08, Gnedin17, Nasir20, Zhu21, Zhu22} and transmission statistics \citep[e.g.][]{Fan06, Becker15, Eilers18, Yang20, Bosman22} as well as \Lya\ damping wing absorption \citep[e.g.][]{DaviesF18, Greig22} suggest that signatures of reionization persist below $z\sim$~6; in agreement with constraints from independent state-of-the-art probes (e.g. the luminosity function of \Lya\ emitters; \citealt{Konno18}, the fraction of Lyman Break galaxies showing \Lya\ emission; \citealt{Mason18}, and the optical depth to reionization; \citealt{Planck18}). 

Precision measurements of absorber properties at $z\sim$~6 offer another independent avenue to probe the reionization history of the Universe. If the CGM is primarily ionized by the UV background (rather than by the radiation field of the host galaxy), then we would expect to see a transition from primarily neutral or low-ionization metal lines while reionization is ongoing to more highly ionized lines once reionization is complete \citep[e.g.][]{Oppenheimer09, Finlator15, Finlator16, Doughty18}. There is growing observational evidence for such a shift of metals towards lower ionization environments at $z\gtrsim$~5. The incidence of high-ionization absorbers such as \mbox{\CIV~$\lambda \lambda$1548, 1550\AA} and \mbox{\SiIV~$\lambda \lambda$1393, 1402\AA} declines smoothly over \mbox{1.5 $\lesssim z \lesssim$ 5} before dropping rapidly at $z\gtrsim$~5 (e.g. \citealt{Sargent88, Songaila01, Boksenberg03, Simcoe06, RyanWeber06, RyanWeber09, Becker09, Simcoe11, DOdorico13, Diaz16, Cooper19, Meyer19, DOdorico22}, Davies et al. submitted), whereas the frequency of weak \OI~$\lambda$1302\AA\ absorbers (tracing neutral gas) increases over this same range \citep[e.g.][]{Becker06, Becker19} and the incidence of \mbox{\MgII~$\lambda \lambda$2796, 2803\AA} absorbers remains approximately constant \citep[e.g.][]{Matejek12, Bosman17, Chen17, Codoreanu17, Zou21}. There is also evidence for an increase in the \CII~$\lambda$1304\AA/CIV\ ratio at $z\gtrsim$~5.7 \citep[e.g.][]{Cooper19, Simcoe20}. All of these results suggest that metals at $z\gtrsim$~5 typically inhabit lower ionization environments than those at lower redshifts. 

Metal absorbers also offer a unique probe of stellar populations at high redshift. The chemical abundance ratios of neutral absorbers reflect the yields of stellar and supernova nucleosynthesis, which are strongly dependent on both the masses and metallicities of the progenitor stars \citep[e.g.][]{Cooke11}. Above some redshift, we expect to observe a shift in the abundance patterns of neutral absorbers to reflect the chemical signatures of metal-free Population III stars \citep[e.g.][]{Simcoe04, Greif07, Welsh22}. Several metal-poor absorption systems have been identified at \mbox{$z >$ 5.5}, but their abundance ratios appear to be consistent with those measured at lower redshifts \citep[e.g.][]{Becker12, DOdorico18, Banados19, Simcoe20}, suggesting that it may be necessary to push observations to even lower column densities and/or higher redshifts in order to uncover the very metal poor systems harbouring the chemical signatures of the first stars in the Universe.

Finally, higher ionization species such as \CIV\ and \SiIV\ probe enriched gas in galaxy halos and are therefore a unique tracer of outflows in the early Universe. The observed incidence of strong \CIV\ absorbers at $z\gtrsim$~5 is significantly larger than predicted by cosmological hydrodynamical simulations, which suggests that either the metal outflow rate in the feedback models is too low or the assumed ionizing photon background does not reflect the true conditions in the early Universe \citep[e.g.][]{Rahmati16, Keating16, Finlator20, DOdorico22}. \citet{Suresh15} find that if the feedback is strengthened to reproduce the observed incidence of \CIV\ absorbers, the galaxy star formation rates are reduced to levels that are inconsistent with measured values. This discrepancy has been further strengthened by the recent observational claims of very massive galaxies at $z\gtrsim$~10 revealed by the \textit{James Webb Space Telescope} (JWST) \citep[e.g.][]{Adams22, Atek22, Furtak22, Harikane22b, Labbe22, Naidu22, Yan22}. This highlights the power of \CIV\ statistics to provide independent constraints on feedback models. 

The ``Ultimate XSHOOTER legacy survey of quasars at $z\sim$~5.8-6.6'' (XQR-30; PI D'Odorico; D'Odorico et al. in prep) has obtained 30 high signal-to-noise (S/N) ($\geq$~10 per 10~\kms\ spectral pixel at 1285\AA\ rest-frame) quasar spectra with the goal of characterizing the properties of the Universe during the second half of reionization. This European Southern Observatory (ESO) large program quadruples the existing sample of 12 archival spectra in the same redshift range observed at comparable S/N and spectral resolution ($\sim$~30~\kms). The combined sample of 42 quasars (hereafter referred to as the ``enlarged XQR-30'' or E-XQR-30 sample) provides the largest homogeneous collection of high quality quasar spectra in the early Universe. 

In this paper we present a catalog of metal absorption systems in the E-XQR-30 spectra. The catalog was constructed using a combination of automated line finding, custom filtering algorithms and visual inspection. The paper is structured as follows. The quasar sample and data processing are described in Section \ref{sec:sample} and the steps taken to construct the absorber catalog are outlined in Section \ref{sec:method}. In Section \ref{sec:absorber_properties} we present the physical properties of the absorber sample and explore how the sample size and absorption search path compare to previous studies in the literature. The completeness and false-positive rate in the catalog are characterized in Section \ref{sec:completeness}. The released products are described in Section \ref{sec:catalog_desc} and the work is summarized in Section \ref{sec:summary}.

Throughout this work we adopt the Planck 2018 $\Lambda$CDM cosmology with \mbox{H$_{0}$ = 67.7 \kms\ Mpc$^{-1}$} and \mbox{$\Omega_m$ = 0.31} \citep{Planck18}. 

\section{Observations and Data Processing}\label{sec:sample}
\subsection{Sample}
The majority of the spectra used in this work were obtained as part of XQR-30, an ESO Large Program (PI: V. D'Odorico) which acquired deep, medium resolution (\mbox{R $\simeq$~10000}, \mbox{FWHM $\simeq$ 30~\kms}) observations of 30 bright \mbox{(J$_{\rm AB} \le 19.8$)} quasars at \mbox{5.8 $\lesssim z \lesssim$~6.6} using XSHOOTER \citep{Vernet11}. The other 12 spectra were sourced from archival XSHOOTER observations of quasars in the same redshift and magnitude range that have similar spectral resolution and S/N to the XQR-30 spectra. The full sample is described in detail in D'Odorico et al. (in prep). All 42 quasars were targeted based on their redshift and magnitude, with no prior knowledge of intervening absorber properties.

\subsection{Data Reduction and Processing}
\subsubsection{Extracting 1D Spectra}
The data reduction process is described in \citet{Becker19}. Briefly, for each exposure, a composite dark frame was subtracted before performing sky subtraction on the un-rectified frame \citep{Kelson03}. An initial 1D spectrum was extracted using the optimal weighting method of \citet{Horne86} and flux calibration was performed using measurements of a standard star. Telluric corrections were computed using the Cerro Paranal Advanced Sky Model \citep{Noll12, Jones13} and applied back to the 2D spectra. A final 1D spectrum was extracted for each of the XSHOOTER spectroscopic arms (VIS and NIR) using all exposures simultaneously to maximize the bad pixel rejection efficiency. The extracted spectra have a velocity sampling of 10~\kms\ per pixel which corresponds to approximately three spectral pixels per resolution element (see Section \ref{subsubsec:spectral_res}). The archival observations were re-reduced using the same method for consistency. 

For each quasar, the VIS and NIR spectra were combined into a single spectrum by scaling the NIR spectrum to match the median value of the VIS arm over \mbox{990~--~1015nm}, truncating both spectra at 1015nm, and stitching them together. We note that the absorption line properties are measured from continuum-normalized spectra (as described in the following section) and therefore the absolute flux scaling of the combined spectrum has no impact on our measurements. In some cases we found small \mbox{(5~--~20~\kms)} offsets between the redshift centroids of transitions of the same ion and absorption system falling in the VIS arm and the NIR arm (e.g. \SiII~$\lambda$1260 vs. \SiII~$\lambda$1526). These offsets are believed to originate from small spatial misalignments between the slits. When such velocity offsets were identified, the NIR spectra (for which the uncertainties on the wavelength solutions are slightly larger) were shifted by $\pm$~5~--~20~\kms\ to align the absorption profiles across both arms. A more detailed description of the velocity shifting and its impact on the relative centroids of the line profiles is given in Appendix \ref{appendix:velshift}.

\subsubsection{Spectral Resolution Measurements}\label{subsubsec:spectral_res}
When fitting the absorption lines it became apparent that some spectra have significantly higher spectral resolution than the nominal value (which depends on the slit width). This occurs when the average seeing is significantly smaller than the slit width (0.9'' in the VIS arm and 0.6'' in the NIR arm for the XQR-30 spectra). We estimated the true velocity resolutions of the final extracted VIS and NIR spectra for each quasar using seeing data for the individual exposures as described in D'Odorico et al. (in prep). Briefly, we used high S/N exposures to measure the relationship between the average telluric line width (a single value per exposure and spectral arm) and the seeing for each slit width and spectral arm, used these relationships to estimate the width of the line spread function for each exposure, created a composite profile weighted by the inverse variance of the individual exposures, and measured the width of the composite profile to obtain the final spectral resolution. 

The adopted spectral resolutions are listed in Table \ref{redshift_table}. The resolutions for the VIS spectra range from \mbox{R = 9500~--~13700} (median 11400, corresponding to \mbox{FWHM = 26~\kms}) and are higher than the nominal resolution of \mbox{$R$ = 8800} \mbox{(FWHM = 34 \kms)} for the 0.9'' slit. The NIR resolutions range from \mbox{$R$ = 7600~--~11000} (median 9750, corresponding to \mbox{FWHM = 31~\kms}) and are comparable to or higher than the nominal resolution of \mbox{R = 7770} \mbox{(FWHM = 39 \kms)} for the 0.6'' slit. The high spectral resolution of our observations allows us to resolve the metal absorption lines more effectively than some of the largest recent surveys of metal absorbers at $z\gtrsim$~5 that have velocity resolutions of \mbox{FWHM $\sim$~50~\kms} (e.g. \citealt{Chen17, Cooper19}; see Section \ref{subsec:absorber_stats}).

\subsubsection{Quasar Redshifts}
We adopted the best quasar redshifts that were available when the construction of the absorption line catalog commenced. The redshifts were taken from emission line measurements where available (primarily [C~\textsc{ii}]~$\lambda$158$\mu$m as well as CO, \MgII, and \Lya~+~\NV), or from the apparent start of the \Lya\ forest (see Table \ref{redshift_table}). The adopted redshifts are consistent with the final emission redshifts published in D'Odorico et al. (in prep) to within $\Delta v \lesssim$~300~\kms. The redshift values could not be updated if better measurements became available because we only search for absorbers redward of the quasar \Lya\ emission line and up to a maximum redshift of 5000~\kms\ above of the quasar redshift (see Section \ref{subsec:system_search}). However, these small redshift differences do not have any significant impact on the final absorber catalog. 

\section{Absorption line catalog}\label{sec:method}
The absorption line catalog was constructed by performing an automated search for candidate systems in each spectrum, filtering the list of candidates using a combination of custom algorithms and visual inspection, and fitting Voigt profiles to obtain the column density ($\log N\equiv \log N$/cm$^{-2}$) and Doppler ($b$) parameter of individual absorption components. Many of the steps use routines from \textsc{astrocook}\footnote{\href{https://github.com/DAS-OATs/astrocook}{\url{https://github.com/DAS-OATs/astrocook}}}, a \textsc{Python} library and application for identifying and fitting absorption systems in quasar spectra \citep{Cupani20}. The following sections provide detailed explanations of all steps.

Throughout the analysis we use only the wavelength region redward of the \Lya\ emission line of each quasar. Neutral hydrogen in the $z\sim$~6 intergalactic medium results in saturation of the \Lya\ forest, making it impossible to identify individual absorption systems at shorter wavelengths.

\begin{figure*}
\centering
 \includegraphics{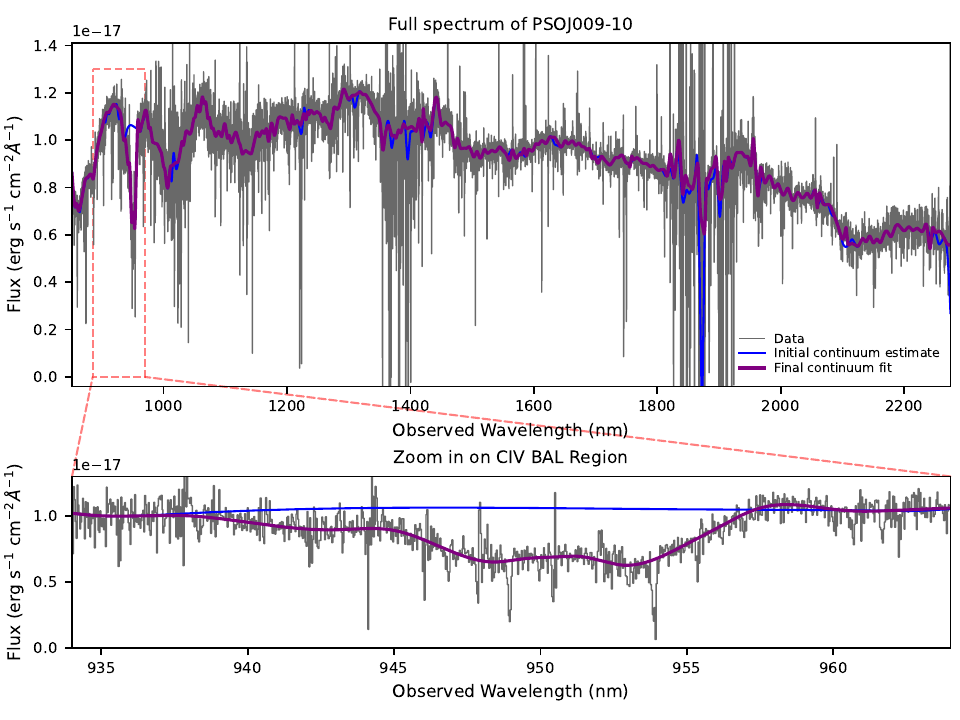}
 \caption{Continuum fit to the spectrum of the BAL quasar PSOJ009-10. The top panel shows the spectrum over the entire range redward of the \Lya\ emission line, and the bottom panel shows a zoom-in around the CIV BAL feature. In both panels the observed spectrum is shown in grey, the initial continuum estimate returned by \textsc{astrocook} is shown in blue, and the purple curve shows the final continuum fit after manual adjustments which were primarily required around the BAL feature and in regions with prominent skyline residuals. }\label{fig:continuum_fit} 
\end{figure*}

\subsection{Continuum Normalization}
A first estimate of the continuum shape is obtained using the \texttt{lines\_find} and \texttt{nodes\_cont} recipes in \textsc{astrocook}. The \texttt{lines\_find} routine creates a list of potential lines by convolving the spectrum with Gaussian profiles of different widths (standard deviation \mbox{0 $< \sigma \leq$ 100~\kms)} and identifying troughs with a depth $>$~5$\times$ the median error. The \texttt{nodes\_cont} recipe masks the identified lines, calculates the median wavelength and flux in windows of fixed velocity width, and inserts nodes at these locations. Finally, univariate spline interpolation is applied to the nodes to estimate the shape of the continuum at the wavelength sampling of the original spectra. We adopt a window width of 770~\kms\ which gives good initial results for our data. 

The initial continuum estimates are manually refined by adding and removing nodes as necessary using the \textsc{astrocook} GUI. Manual refinement is most commonly required in regions where:
\begin{itemize}
 \item Strong skyline or telluric residuals cause the measured flux to be highly variable, resulting in large oscillations in the estimated continuum level.
 \item Clustered narrow absorption lines fill a significant fraction of the 770~\kms\ window, leaving relatively few unmasked channels to estimate the median flux level.
 \item The spectrum has a significant slope which is not well captured at the adopted 770 km/s sampling.
 \item There is a prominent broad absorption line (BAL) feature.
\end{itemize}

A significant fraction of the E-XQR-30 spectra show BAL features (\citealt{Bischetti22} and submitted). When processing the spectra of BAL quasars, we attempt to adjust the continuum fit to follow the BAL troughs because this increases the probability of recovering underlying narrow absorption lines. Figure \ref{fig:continuum_fit} shows the initial continuum estimate and final continuum fit for the BAL quasar PSOJ009-10. The top panel shows the fits over the full spectral range redward of \Lya\ and the bottom panel shows a zoom-in around the \CIV\ BAL feature, where the continuum fit has been manually adjusted to follow the BAL trough. Absorbers found within BAL regions are flagged in the final catalog (see Section \ref{subsec:flags} and Column 14 in Table \ref{table:catalog}). The final continuum fits are made publicly available with this paper (see Section \ref{sec:catalog_desc}). 

\begin{table*}
\begin{nscenter}
\begin{tabular}{l|c|c|c|c|c|c|c}
\hline
 &  &  & \multicolumn{2}{c|}{$\Delta X$ (Intervening)} & \multicolumn{2}{c|}{$\Delta X$ (all)} & \\ 
Ion & Transitions & $z$ range & XQR-30 & E-XQR-30 & XQR-30 & E-XQR-30 & Systems \\ \hline
\MgII\ & \MgII~$\lambda$2796 \& $\lambda$2803 & 1.944~--~6.760 (6.381) & 400.7 & 553.3 & 445.0  & 612.4 & 360\\ \hline 
\FeII\ & Pairs of \FeII~$\lambda$2344, $\lambda$2382,  & 2.183~--~6.759 (6.380) & 422.8 & 594.3 & 473.3  & 665.1 & 184\\ 
 & $\lambda$2586, $\lambda$2600 &  &  &  & & & \\ \hline
\CIV\ & \CIV~$\lambda$1548 \& $\lambda$1550 & 4.317~--~6.760 (6.339) & 145.6 & 211.8 & 193.8  & 280.3 & 479\\ \hline 
\SiIV\ & \SiIV~$\lambda$1393 \& $\lambda$1402 & 4.907~--~6.715 (6.339) & 83.6 & 120.6 & 126.0  & 182.5 & 127\\ \hline 
\CII\ & \CII~$\lambda$1334 paired with \OI~$\lambda$1302,  & 5.169~--~6.760 (6.381) & 55.4 & 77.8 & 105.9  & 148.6 & 46\\ 
 & \AlII~$\lambda$1670, \SiII~$\lambda$1260, $\lambda$1526 &  &  &  & & & \\ \hline
\NV\ & \NV~$\lambda$1238 \& $\lambda$1242 & 5.645~--~6.760 (N/A) &  0.0 &  0.0 & 33.1  & 47.6 & 13\\ \hline
\end{tabular}
\end{nscenter}
\caption{Transitions and redshift ranges used when searching for candidate absorption systems, unmasked total absorption path lengths (see Section \ref{subsec:abs_path}) and system detection statistics. Candidate absorption systems are identified by searching for pairs of transitions with $\geq$~98\% probability of absorption at the same redshift, as described in Section \ref{subsec:system_search}. Candidate \CII\ systems are rejected if either \OI~$\lambda$1302 or \SiII~$\lambda$1260 fall within the probed wavelength range but are not detected. The bracketed values in column 3 indicate the maximum search redshift for intervening absorbers. We note that the maximum search redshifts differ between ions because the search paths exclude regions that are masked due to skyline contamination or poorly corrected telluric absorption, as well as regions showing broad absorption line (BAL) features (see Section \ref{subsec:abs_path}). $\Delta X$(intervening) gives the path length for intervening absorbers alone (separated by at least 10,000~\kms\ from the quasar redshift), whereas $\Delta X$(all) also includes the search path for proximate absorbers. We note that \FeII\ is accessible over a smaller $z$ range than \MgII\ but has a larger $\Delta X$ because \FeII\ systems can be detected in any two of the four listed transitions, meaning that masking removes less of the search path.}\label{linetable}
\end{table*}

\subsection{Identification of Candidate Metal Absorption Systems}\label{subsec:system_search}
After dividing out the continuum, we search for candidate absorption systems in six of the most commonly observed ions: \MgII, \FeII, \CII\ (combined with \OI\ and \SiII~$\lambda$1260\AA\ when accessible), \CIV, \SiIV, and \NV\ (specific transitions listed in Table \ref{linetable}). Of these, \FeII, \CII, \SiII\ and \OI\ trace low-ionization gas, \CIV, \SiIV\ and \NV\ trace high-ionization gas, and \MgII\ traces a range of ionization states. Searching for a combination of ions spanning a wide range of ionization energies enables many interesting science cases. For example, the \CIV/\SiIV\ ratio is sensitive to the shape of the ionizing photon background \citep[e.g.][]{Songaila95, Songaila98, Bolton11, Finlator16, Doughty18, Graziani19}, and the \CIV/\CII\ ratio directly probes the distribution of carbon between ionization states \citep[see e.g.][]{Prochaska15, Cooper19, Berg21}.

We construct a list of candidate absorption systems in each ion by searching for coincident absorption in pairs of transitions. For the ions observed as doublets (\MgII, \CIV, \SiIV\ and \NV), we search for matching absorption in the two transitions of the doublet. For \FeII, we search for matching absorption in any pair of the four strongest transitions ($\lambda$2344, 2382, 2586 and 2600\AA). This increases the probability of detection compared to searching for only the two strongest lines because one or more transitions may fall in a region contaminated by skyline or telluric residuals. There is only one strong \CII\ transition in the probed spectral region, so it is necessary to search for associated low-ionization transitions of other ions to confirm the redshift. We search for \CII~$\lambda$1334 paired with \OI~$\lambda$1302, \SiII~$\lambda$1260, \SiII~$\lambda$1526, or \AlII~$\lambda$1670. If either of \SiII~$\lambda$1260 or \OI~$\lambda$1302 fall within the probed wavelength range but are not detected, the system is rejected because \CII~$\lambda$1334, \OI~$\lambda$1302 and \SiII~$\lambda$1260 are expected to have comparable equivalent widths \citep[e.g.][]{Becker19}. \AlII~$\lambda$~1670 and \SiII~$\lambda$1526 are not required but enable the detection of lower redshift systems for which \SiII~$\lambda$1260 and/or \OI~$\lambda$1302 fall outside of the probed wavelength range. The considered transitions and redshift search interval for each ion are listed in Table \ref{linetable}. At this stage we do not search for systems that show \OI\ or \SiII\ absorption without associated \CII\ absorption. We later search for additional transitions associated with all candidate absorption systems (see Section \ref{subsec:systs_complete}), meaning that \OI\ and \SiII\ lines could be detected as associated transitions of \MgII\ or \FeII\ systems. However, we note that all \OI\ and \SiII~$\lambda$1260 systems in our final catalog have associated \CII\ absorption, and we recover all \OI\ systems reported by previous surveys (see Section \ref{subsec:cross_check}).

For each line pair, the lower end of the redshift search interval is fixed by the wavelength of the quasar \Lya\ line and the maximum search redshift is set to 5000~\kms\ above the quasar redshift\footnote{Our final catalog includes seven absorption systems that lie above the redshift of the relevant quasar. In five of these cases the quasars have redshifts measured from \Lya\ emission/absorption, which are less accurate than [C~\textsc{ii}] or CO redshifts.}. \NV\ is produced by photons with energies $\geq$~77.5~eV and is therefore expected to be found in systems illuminated by the radiation field of a quasar \citep[e.g.][]{Perrotta16}. Because of the small wavelength separation between \NV~$\lambda$1238 and \Lya~$\lambda$1216, the search interval for \NV\ is restricted to `proximate' systems with velocity separations between $-$5460~\kms\ and +5000~\kms\ from the quasar redshift (see Section \ref{subsec:proximate_intervening}). 

The search for candidate systems is performed using the \textsc{astrocook} recipe \texttt{systs\_new\_from\_like}. The redshift search interval is divided into bins of a specified width (we adopted \mbox{$\Delta z$ = 10$^{-4}$} corresponding to \mbox{$\Delta v \simeq$~4~--~10~\kms}). The quasar spectrum is then resampled onto this redshift grid for both transitions of the relevant line pair. For each redshift bin, the absorption S/N (defined as \mbox{(1$-$flux)/error}) is calculated for each transition and the Gaussian Error Function is used to determine the probability that a feature with this S/N does not trace noise in the spectrum, evaluated as \mbox{$P_i$ = \textsc{erf}([S/N]/$\sqrt{2}$)}. The absorption probabilities for the two transitions are combined using an empirical expression\footnote{\mbox{$P = 1 - (1 - P_1 \times P_2)^2$}. The absorption probabilities for the individual transitions are multiplied, the product is subtracted from 1 which maps peaks to valleys, the expression is squared to enhance extremes and again subtracted from 1 to map the valleys back to peaks. We note that in future versions of \textsc{astrocook} this expression will be replaced with ERF$^{-1}(P_1 \times P_2)$.} to estimate the probability that both transitions show absorption at that redshift. Finally, a peak-finding algorithm is used to identify peaks in redshift space that exceed a given threshold probability and are separated by a specified minimum number of redshift bins. We tested a range of parameters and adopted a threshold probability of $P >$ 0.98 and a minimum peak separation of 4 bins because these values were found to maximize the recovery rate of visually identifiable absorbers whilst minimizing the number of obviously spurious candidates (e.g. those associated with skyline residuals). 

When performing this search, we manually mask wavelength regions with particularly bad skyline or telluric residuals (primarily in the gaps between the $J$, $H$ and $K$ NIR bands at \mbox{$\lambda \simeq$~1345~--~1450nm} and \mbox{1800~--~1940nm}). Including these regions would result in a large number of spurious candidate systems. However, individual skylines in other spectral regions are left unmasked. Some spectra show strong absorption features close in wavelength to the quasar \Lya\ emission line, making it difficult to determine the strength of the overlying continuum. In such cases we mask the affected regions which can extend up to $\sim$~10\AA\ (observed) redward of the \Lya\ emission. The masked regions are listed in Table \ref{masked_region_table}.

\subsection{Automatic Filtering of Candidate List}\label{subsec:auto_filtering}
\subsubsection{Overview}
The initial line search outputs six files, each of which records the properties of candidate absorption systems detected in one of the six primary ions. Each entry of each file contains the redshift of one candidate system and the pair of transitions in which the absorption was detected (e.g. 6.1459, [\OI~$\lambda$1302, \CII~$\lambda$1334]). These individual file entries are hereafter referred to as `absorption pairs'. 

More than 50\% of the initial candidates are likely to be false-positives (see Section \ref{subsec:false_positive}). We therefore use a custom filtering algorithm to decrease the amount of contamination in the sample prior to further processing. The filtering algorithm specifically targets spurious absorption systems with two main origins:
\begin{itemize}
 \item Negative skyline residuals falsely identified as absorption lines (see example 1 in Figure \ref{fig:filtering}). Candidate systems are classified as spurious if one or both absorption profiles span less than three 10~\kms\ spectral channels (approximately one resolution element).
 \item Chance alignment of unassociated absorption lines. This occurs when two absorption lines fall at approximately the velocity separation expected for a given line pair, but actually trace absorption in other ions at different redshifts. Such systems are identified by examining whether the absorption profiles of the two transitions are aligned in velocity space (example 2 in Figure \ref{fig:filtering}) and, for transitions of the same ion, whether their equivalent width ratio is consistent with expectations from quantum mechanics (example 3 in Figure \ref{fig:filtering}).
\end{itemize}

The filtering algorithm is applied to each absorption pair individually. We note that the rejection of any given pair does not preclude the acceptance of other absorption pairs at the same redshift. As described in Section \ref{subsec:system_search}, the absorption pairs mostly trace two transitions of the same ion, except in the case of \CII~$\lambda$1334 where the accompanying line is one of \OI~$\lambda$1302, \AlII~$\lambda$1670, \SiII~$\lambda$1260 or \SiII~$\lambda$1526.

The automatic filtering reduces the number of candidate absorption systems by $\sim$~35~--~50\% depending on the ion. It is not possible to determine what percentage of the rejected systems are real. Based on the mock absorber catalog used for the completeness calculations (see Section \ref{subsec:mock_spectra}), we estimate that this filtering only reduces the sample completeness by a few percent while lowering the false-positive rate by $\sim$~20\%, from an initial value of $\sim$~70\% to a post-filtering value of $\sim$~50\% (see Section \ref{subsec:false_positive}).

\subsubsection{Spurious Candidates Associated with Skyline Residuals}\label{subsubsec:skylines}
A significant fraction of the spurious candidates arise due to unmasked negative skyline residuals. The top panel of Figure \ref{fig:filtering} shows an example of such a candidate. Each row shows the absorption profile for one transition of the absorption pair (labelled Line 1 and Line 2), resampled onto a velocity grid centered at the expected absorption wavelength. In this candidate, Line 1 traces real absorption but Line 2 is clearly a skyline residual. We minimize contamination from skyline residuals by rejecting candidates for which the putative absorption does not span at least three 10~\kms\ velocity channels (approximately one resolution element) in both transitions. We emphasize that this filtering does not preclude Line 1 from being accepted as part of a different absorption pair.

\begin{figure}\label{fig:filtering}
\centering
 \includegraphics[scale=0.55, clip = True, trim = 10 90 0 10]{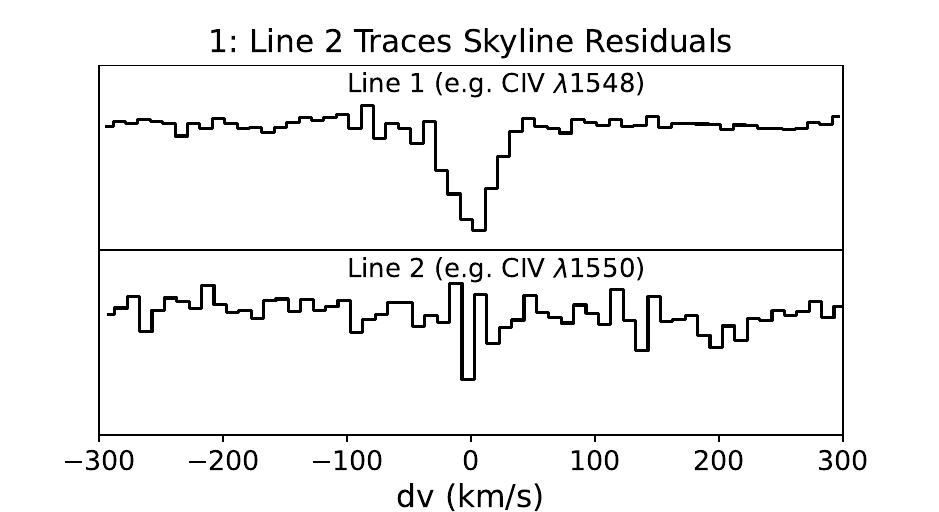} \\
 \includegraphics[scale=0.55, clip = True, trim = 10 0 0 10]{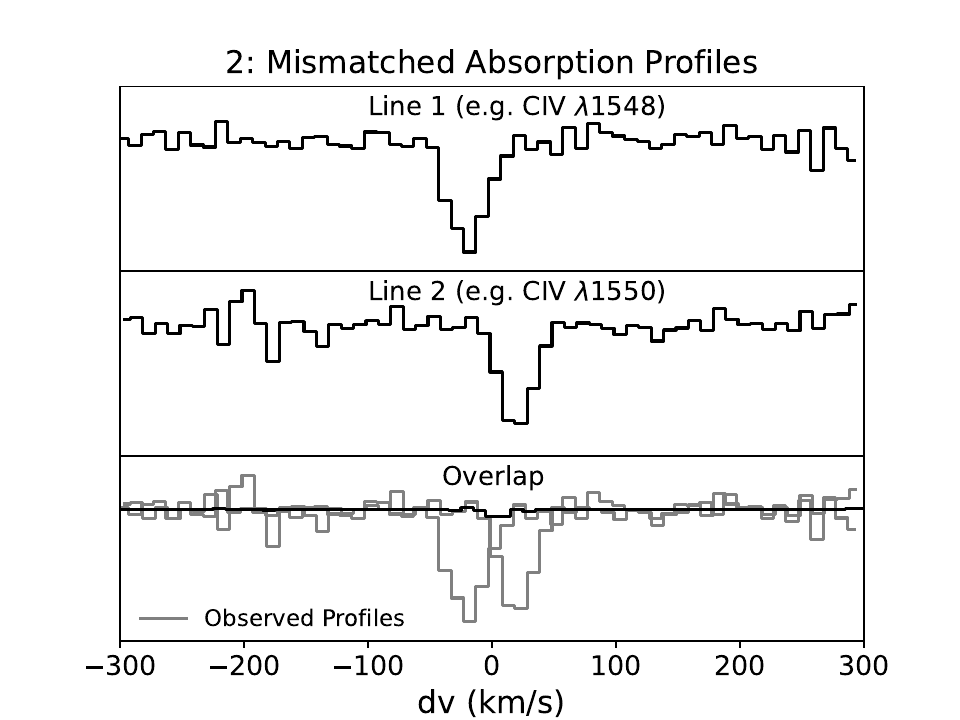} \\
 \includegraphics[scale=0.55, clip = True, trim = 10 90 0 10]{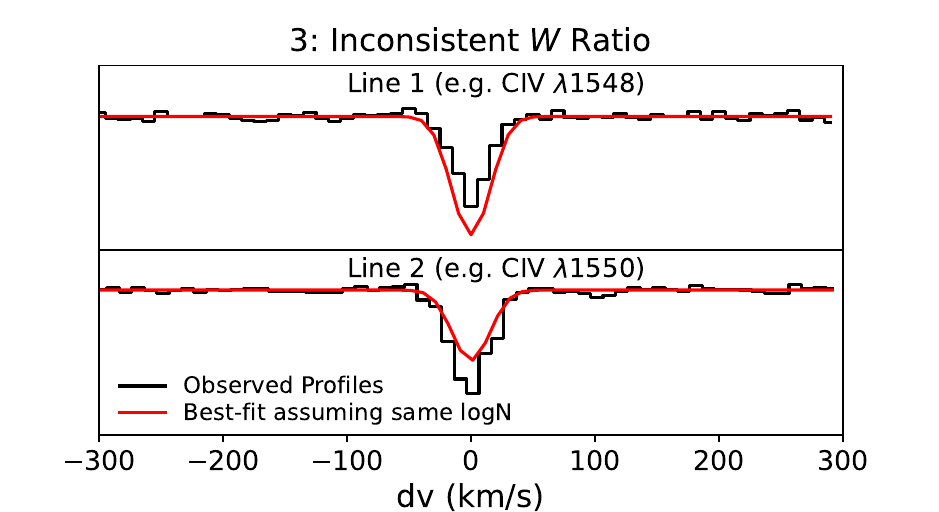}
 \caption{Illustration of the criteria used to filter the lists of candidate absorption pairs. The absorption profile of each transition is resampled onto a 10~\kms\ grid centered at the redshift of the candidate. \textit{Top}: Line 2 is a skyline residual. A candidate is rejected if either absorption line does not span at least three 10~\kms\ velocity channels (about one resolution element; Section \ref{subsubsec:skylines}). \textit{Middle and Bottom}: In each candidate, both transitions are real absorption lines but they are associated with unrelated absorption systems. In the middle panel, the absorption profiles are misaligned in velocity space. We reject candidates that do not have significant absorption overlapping between the two transitions (Section \ref{subsubsec:overlap_signal}). The candidate in the bottom panel is rejected because the observed equivalent widths (black) are inconsistent with Voigt profile fits assuming a single column density (red) (Section \ref{subsubsec:ew_test}). } \label{fig:filtering} 
\end{figure}

\subsubsection{Velocity Alignment of Absorption Profiles}\label{subsubsec:overlap_signal}
The majority of the remaining spurious candidates arise when two real absorption lines are observed at a velocity separation similar to that expected for the relevant absorption pair, but they actually trace other transitions at different redshifts. For example, \OI~$\lambda$1302 and \SiII~$\lambda$1304 are separated by 507~\kms\ and may therefore masquerade as \CIV~$\lambda\lambda$1548,1550 which has a separation of 499~\kms. Such cases are identified in two ways. Firstly, we examine whether the absorption profiles of the two transitions are aligned in velocity space by computing the overlap signal:
\begin{equation}
 {\rm Overlap}(v) = 1 - (1 - {\rm Line \, 1}(v)) \times (1 - {\rm Line \, 2}(v))
\end{equation}
The middle panel of Figure \ref{fig:filtering} shows the absorption profiles of a candidate absorption pair (top and middle rows) and the overlap signal between them (bottom row). The two absorption profiles are offset in velocity space, and as a result there is no significant overlap signal. 

We use \textsc{Astrocook} to fit a Voigt profile to the overlap signal for each candidate absorption pair (see Section \ref{subsec:voigt_fitting}). An absorption pair is only rejected if the best-fit column density is equal to the lower bound adopted in the fitting ($\log (N$/cm$^{-2})$ = 10), because this indicates that the measured $\log N$ is an upper limit and therefore that there is no significant overlap between the absorption profiles. We emphasize that this criterion does not require that the two absorption profiles have exactly the same velocity structure, it only requires that there is some overlapping absorption. This is to prevent the spurious rejection of systems that have multiple kinematic components and/or are blended with unassociated absorption lines at different redshifts. The absorption pairs trace either two transitions of the same ion which are expected to have the same kinematics, or in the case of pairs detected using \CII, two low-ionization transitions that are expected to have very similar kinematics (see Section \ref{subsec:voigt_fitting}).

\subsubsection{Equivalent Width Ratio Test}\label{subsubsec:ew_test}
For candidate absorption pairs where Line 1 and Line 2 are transitions of the same ion (i.e. pairs detected in all ions \textit{except} \CII), we also check whether the relative equivalent widths ($W$) of the two absorption profiles are consistent with expectations from quantum mechanics. We fit Voigt profiles to the two transitions assuming that they have the same $z$, $b$ parameter, and $\log N$, measure the equivalent widths of these best-fit profiles, and use these to calculate the `expected' ratio and difference between the equivalent widths of the stronger and weaker transition ($W_{\rm strong}/W_{\rm weak}$ and $W_{\rm strong} - W_{\rm weak}$). By computing the expected parameters in this way we naturally account for changes in the relative equivalent width as the absorption optical depth increases. We then repeat the Voigt profiles fits, tying the $z$ and $b$ parameter as before but this time allowing the two transitions to have different $\log N$ values. This second set of Voigt profiles is then used to calculate the `measured' $W$ ratio and difference. We reject pairs for which the measured $W$ ratio and/or difference are inconsistent with the expected values at the $>$~1$\sigma$ level (see Section \ref{subsec:ew_calculations} for a description of the $W$ error calculations). The bottom panel of Figure \ref{fig:filtering} shows an example of a candidate absorption pair for which the equivalent widths of the observed profiles (black) are inconsistent with expectations for absorption at a single column density (red). 

If enforced uniformly, this criterion could incorrectly reject systems for which one or both of the transitions is blended with unassociated strong absorption at a different redshift. When multiple transitions fall within the full width at ten percent maximum (FWTM) of a given absorption profile, we conservatively treat the expected equivalent width calculated for that profile as a lower limit. This also applies to absorption profiles with multiple kinematic components. A multi-component absorption profile may be identified as a single candidate system (e.g. when there is one dominant component accompanied by other weaker components, such as in rows 2 and 3 of Figure \ref{fig:multi_comp_fitting}) or may be split into multiple candidates (e.g. when the absorption profile shows multiple clear peaks as seen in Figure \ref{fig:fitPlots}). In the first scenario, the measured $W$ ratio reflects the relative equivalent widths of the dominant component. In the second scenario, the $W$ ratio test is applied to each component separately. As before, if any component is within the FWTM of another component, the expected $W$ is treated as an upper limit. 

The equivalent width check cannot be applied to \CII\ pairs because we probe only one \CII\ transition. To minimize the fraction of spurious \CII\ systems we instead require candidates to be detected in at least three low-ionization transitions (see Section \ref{subsec:systs_complete}).

\subsubsection{Removal of Duplicates}
Noise in the spectrum can cause the peak-finding algorithm to identify a single absorption component multiple times. To minimize the number of duplicates in our candidate list we remove candidate systems of the same ion with redshift separation \mbox{$|\Delta z| <$ 5~$\times$~10$^{-4}$} (corresponding to velocity separations \mbox{$\lesssim$~20~--~50~\kms}; approximately one resolution element). 

\subsection{Identification of Candidate Associated Transitions}\label{subsec:systs_complete}
The next step is to search for additional transitions\footnote{We consider all of the transitions listed in Table \ref{linetable} as well as \FeII~$\lambda$1608, \FeII~$\lambda$2374, \CII*~$\lambda$1335, \SiII~$\lambda$1304, \SiII~$\lambda$1808, \SiII*~$\lambda$1264, \MgI~$\lambda$2852, \AlIII~$\lambda$1854, \AlIII~$\lambda$1862, \MnII~$\lambda$2576, \MnII~$\lambda$2594, \MnII~$\lambda$2606, \SII~$\lambda$1250, \SII~$\lambda$1253, \SII~$\lambda$1259, \CoII~$\lambda$1466, \CoII~$\lambda$1574, \CoII~$\lambda$1941, \CoII~$\lambda$2012, \CrII~$\lambda$2056, \CrII~$\lambda$2062, \CrII~$\lambda$2066, \CaII~$\lambda$3934, \CaII~$\lambda$3969, \NaI~$\lambda$5891, and \NaI~$\lambda$5897.} associated with each candidate absorption system. We do this using the \textsc{Astrocook} recipe \texttt{systs\_complete\_from\_like} which employs the same algorithm used to identify candidate absorption systems in Section \ref{subsec:system_search}, but this time calculates the probability that an additional transition is present at the redshift of a given system. When peforming this search, the wavelength masks used in Section \ref{subsec:system_search} are removed to enable the identification of associated transitions lying in noisy spectral regions. Searching for transitions at the redshifts of previously identified systems significantly increases the probability that the recovered absorption is real compared to a blind search. We only accept associated transitions for which the absorption significance exceeds 2$\sigma$.

At the conclusion of this step, we reject any \CII\ systems that are not detected in at least two other strong low-ionization transitions\footnote{\OI~$\lambda$1302, \SiII~$\lambda$1260, \SiII~$\lambda$1526, \AlII~$\lambda$1670, \FeII~$\lambda$2382, \FeII~$\lambda$2600, \MgII~$\lambda$2796, and \MgII~$\lambda$2803} (this removes $\sim$~20\% of the candidates). We do this because the equivalent width ratio test described in Section \ref{subsec:auto_filtering} cannot be applied to \CII, so the probability of spurious candidates resulting from the chance alignment of unassociated transitions in velocity space is elevated compared to the other ions. Requiring the detection of three transitions (including \CII~$\lambda$1334) significantly reduces the number of spurious candidate \CII\ systems.
    
\subsection{Visual Inspection}\label{subsec:visual_inspection}
The remaining candidate systems were visually inspected by five members of the XQR-30 collaboration independently. For each ion in each absorption system, the checkers were asked to indicate whether they thought any real absorption was present. This step is necessary, even for systems with candidate absorption detected in several transitions, because the associated transitions found in the previous step could trace skyline residuals or unrelated absorption lines at other redshifts as discussed in Section \ref{subsec:auto_filtering}. Any absorption classified as spurious by at least three of the five checkers was removed. This visual inspection further reduces the number of candidate absorption systems by \mbox{$\sim$~50~--~75\%} depending on the ion. 

\subsection{Voigt Profile Fitting}\label{subsec:voigt_fitting}
Finally, we use the \textsc{astrocook} GUI to fit Voigt profiles and obtain the column density and $b$ parameter for each component of each absorption system. At this stage there are a total of 964 candidate absorbers of which 505 are detected in \MgII, 262 in \FeII, 263 in \CII, 727 in \CIV, 182 in \SiIV\ and 31 in \NV. The candidate list indicates which transitions show absorption at which redshifts, but it was often necessary to add multiple Voigt profile components per candidate in order to reproduce the observed velocity structure of the absorption profiles.

The Voigt parameters are allowed to vary in the ranges \mbox{$b$ = 5~--~200~\kms} and \mbox{$N$ = 10$^{10}$~--~10$^{18}$~cm$^{-2}$}. If the best-fit $b$ parameter of any component is below the minimum measurable value\footnote{The minimum measurable value is $\sim$~1/3 of the $b$ parameter corresponding to the resolution element, $b_{\rm min} \simeq$~$\frac{1}{3} \frac{c}{R \times 2\sqrt{\ln 2}}$}, we fix it to that minimum value. The reported uncertainties on $\log N$ and $b$ are the 1$\sigma$ statistical errors returned by the \textsc{lmfit} least squares minimization routine used within \textsc{astrocook}.

\begin{figure}
\centering
 \includegraphics[scale=0.55, clip = True, trim = 0 10 0 0]{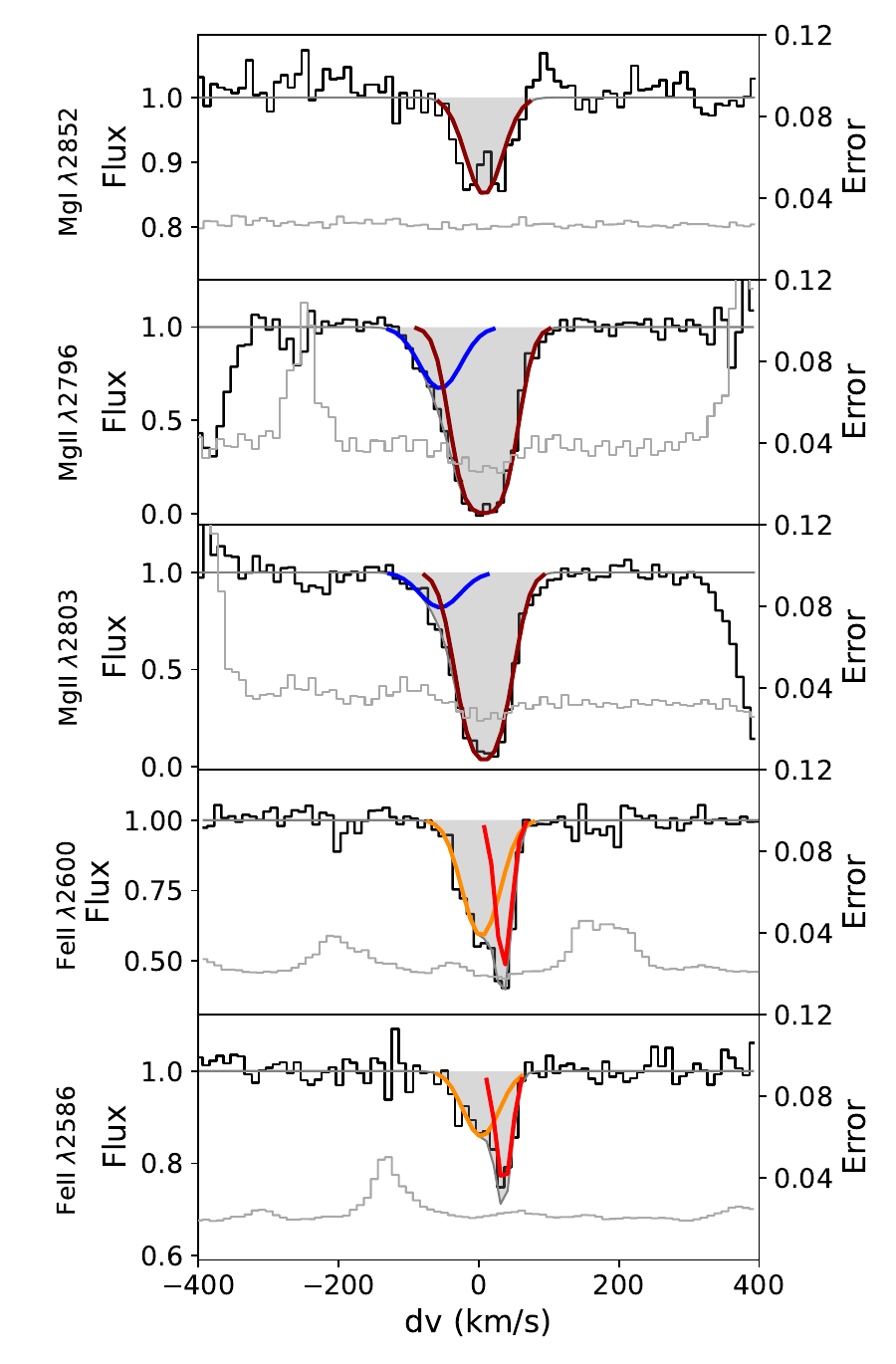}
 \caption{Four component fit to a low-ionization absorption system at \mbox{$z$ = 2.72}. Each color indicates a different kinematic component. The \MgII\ profiles (rows 2 and 3) show a blueshifted wing that is not detected in the other ions. The \FeII\ profiles (rows 4 and 5) are well fit by two components (orange and red). However, the corresponding central absorption trough in \MgII\ (rows 2 and 3) can only be fit with a single component (brown) due to saturation. The \MgI\ absorption profile (row 1) is fit using the same central component (brown) because the low S/N prevents a robust multi-component decomposition.} \label{fig:multi_comp_fitting} 
\end{figure}

Where possible, the kinematics of all low-ionization (neutral or singly ionized) transitions and the kinematics of all high-ionization (doubly ionized or higher) transitions are tied within each component\footnote{The version of \textsc{astrocook} used in this paper (v1.0.0) ties the $b$ parameter by fitting a single value for all transitions, which is equivalent to assuming that the turbulent component is dominant.}. Many systems show both low-ionization and high-ionization transitions, and the transitions in each ionization group are fit separately. However, for $\sim$~15\% of absorption systems it is not possible to fit all ions from one or both ionization categories using the same kinematic structure\footnote{Of these 15\%, $\sim$~80\% are low-ionization systems and $\sim$~20\% are high-ionization.}. Some rare systems show evidence of intrinsic differences between the kinematics of different low-ionization ions, but in most cases the main limitations are saturation and S/N, as illustrated in Figure \ref{fig:multi_comp_fitting}. In this low-ionization absorption system at \mbox{$z$ = 2.72}, the \MgII\ absorption profile (rows 2 and 3) shows a blue-shifted wing which is not detected in either \FeII\ or \MgI. \textsc{Astrocook} does not provide upper limits on column densities of undetected components, so if the blue wing is fit to the \FeII\ and \MgI\ profiles, the best-fit $\log N$ tends towards the minimum allowed value in both cases. Therefore, we remove these components.

In this system, the absorption at \mbox{$\Delta v \simeq$~0} is also fit with a differing number of components depending on the ion. The \FeII\ absorption profile is well fit by two components (shown as orange and red curves). However, the \MgII\ absorption profiles are saturated, and attempting to fit them using three components (the blue wing plus the two \FeII\ components) results in large errors on the column densities of the individual components, indicating that the multi-component decomposition is poorly constrained. Therefore, the \MgII\ profile is fit using two components: the blue wing, and a component encompassing the whole central absorption trough (brown curve). The \MgI\ absorption is relatively weak and the low S/N prevents a robust multi-component decomposition, so it is fit using the same single central component. 

We emphasize that the kinematics are always tied when like-ionization transitions are fit using the same kinematic structure (same number of components at the same velocities). Differing kinematics are only used in cases where it is not possible to obtain well constrained fits for all low-ionization or high-ionization ions with a single set of kinematic parameters.

We find six systems for which different transitions of the same ion cannot be fit using the same $\log N$, perhaps due to partial coverage or errors in continuum normalization. These systems were not rejected by the equivalent width filtering criterion either because the discrepancy was smaller than the associated uncertainty or because there were multiple candidate absorption systems contributing to the absorption profile(s) (see Section \ref{subsec:auto_filtering}). For these systems, we fit each transition independently and the $\log N$ and d$\log N$ given in the catalog represent the average and half-range of the individual $\log N$ values, respectively. This results in larger errors which more accurately reflect the uncertainty compared to the errors returned by \textsc{astrocook}. The $\log N$ values for these systems are flagged (with `M' for multiple in Column 10; see Table \ref{table:catalog}) to indicate that they are a combination of multiple measurements. The reported equivalent widths are measured from the best-fit Voigt profile for each individual line. 

After completing the fitting we perform an automated search for unidentified absorption features in the quasar spectra. Specifically, we searched for regions where the measured flux was significantly below the best continuum + line model obtained from the manual Voigt profile fitting. In some cases, unidentified absorption was attributed to less common transitions that were not part of the initial linelist but were clearly detected in a few systems (e.g. \ZnII\ and \TiII). In other cases, the absorption could be matched to candidate systems that were rejected at the visual inspection phase due to low S/N and/or skyline contamination. In such cases the rejected systems were added back as the best explanation for the observed absorption. The final catalog includes flags to indicate whether each system was automatically or manually identified, and whether or not it passed the visual inspection phase (see Section \ref{subsec:flags} and Column 14 of Table \ref{table:catalog}).

The numbers of systems that were manually added and removed during the fitting stage were sufficiently small that the total number of candidate systems was approximately unchanged.

\subsection{Grouping into Systems}\label{subsec:system_grouping}
We group the components into systems following a method similar to that described in \citet{DOdorico22}. For each quasar we iterate through the list of systems and combine components that are separated by less than 200~\kms\ into a single system characterized by the total equivalent width and equivalent-width-weighted average redshift of the constituent components. This process is repeated until no components are separated by less than 200~\kms. The grouping is applied to all transitions of all ions (both low and high ionization) simultaneously, meaning that a single system can contain both low and high ionization transitions.

\begin{figure}
\centering
 \includegraphics[scale=0.55, clip = True, trim = 10 10 0 0]{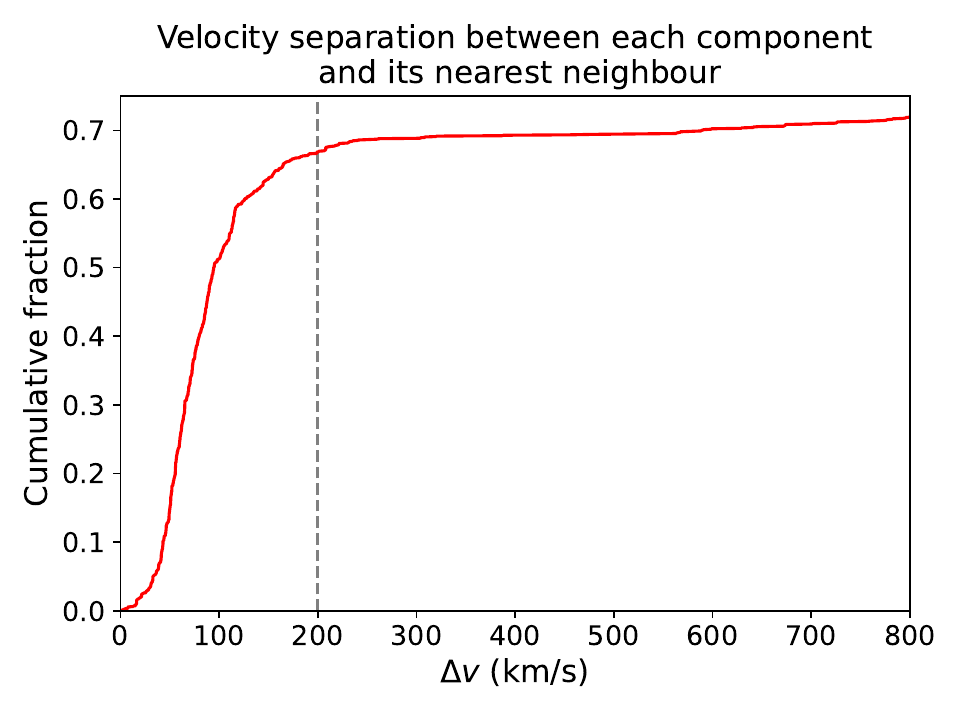}
 \caption{Cumulative distribution of the velocity separation between each absorption component and the closest neighbouring component of the same transition. The cumulative fraction increases rapidly up to a velocity separation of 200~\kms\ and then plateaus, indicating that 200~\kms\ is a physically motivated threshold to adopt when grouping components into systems.} \label{fig:component_separation} 
\end{figure}

The 200~\kms\ threshold was chosen by examining the distribution of velocity separations between each absorption component and the nearest neighbouring component of the same transition, shown in Figure \ref{fig:component_separation}. The cumulative fraction increases rapidly up to a velocity separation of 200~\kms\ and then plateaus, indicating that the vast majority of physically associated components have velocity separations less than 200~\kms. We note that the catalog contains all the information required to generate the line profiles and implement a different grouping method if needed.

\subsection{Calculating Equivalent Width and $\Delta v_{90}$}\label{subsec:ew_calculations}
We measure the rest-frame equivalent width ($W$) of each transition in each system using the Voigt profile fits. This avoids errors associated with skyline spikes and blending with absorption systems at other redshifts. The error on the equivalent width is computed by summing the flux error in quadrature over the region enclosing 90\% of the absorption optical depth. 

$\Delta v_{90}$ is defined as the velocity interval enclosing 90\% of the absorption optical depth \citep{Prochaska97}. We measure $\Delta v_{90}$ using the apparent optical depth ($\tau_a$) profile of the strongest transition of each ion in each system. We construct the cumulative distribution function (CDF) for $\tau_a$ as a function of velocity, and calculate the difference between the velocities at which the CDF equals 0.95 and 0.05.  

\subsection{Classifying Absorbers as Intervening or Proximate}\label{subsec:proximate_intervening}
Absorbers are typically classified as either intervening or proximate. Proximate (also known as `associated' or `intrinsic') absorbers are traditionally defined to lie within $\sim$~5000~\kms\ of the quasar redshift. Due to their close proximity to the quasar, such absorbers may have different ionization states or metal abundance patterns to the underlying absorber population \citep[e.g.][]{Foltz86,Berg16,Perrotta16}. For this reason, it is important to distinguish between intervening and proximate absorbers when investigating e.g. the evolution of absorber properties over cosmic time.

\citet{Perrotta16} found an 8$\sigma$ excess of \CIV\ absorbers (compared to the average incidence) at velocity separations extending down to $-$10,000~\kms\ from the quasar redshift (see also \citealt{Wild08}). The excess at the largest absolute velocity offsets (\mbox{$-$10,000~\kms} to \mbox{$-$5,000 \kms}) is dominated by weak \CIV\ systems (with equivalent width \mbox{$W_{1548} <$ 0.2\AA}), suggesting that the velocity threshold for proximate absorbers should be extended when weak systems are included.  Our absorber sample is 90\% complete at \mbox{$W >$ 0.1\AA} (see Section \ref{subsec:completeness}) and most of the \CIV\ absorbers in our sample are in the \mbox{$W_{1548} <$ 0.2\AA} regime (see Section \ref{subsec:absorber_stats}). Therefore, we classify all systems that have velocity offsets between $-$10,000~\kms\ and +5000~\kms\ from the quasar redshift as proximate absorbers. However, we note that the data published with this paper enable users to calculate the velocity offset of each absorber and apply a different velocity threshold if desired.

\begin{figure*}
\centering
 \includegraphics[scale=1]{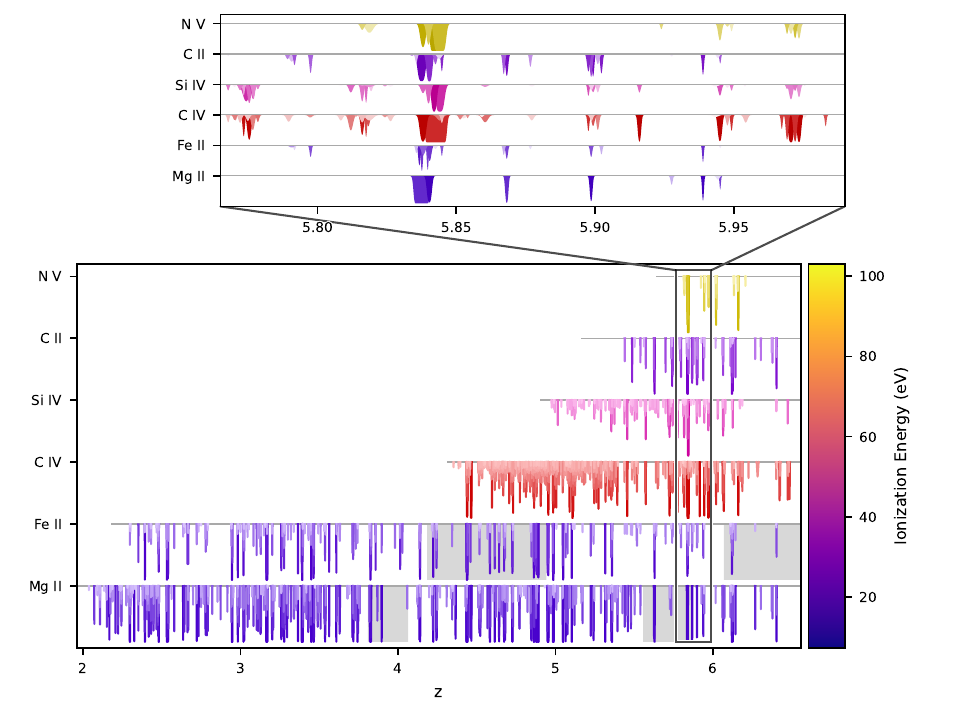}
\vspace{-10pt}
 \caption{Visualisation of the absorber catalog presented in this paper. In the bottom panel, each row shows all of the absorption systems detected in one of the six primary ions across all of the E-XQR-30 spectra combined. For each system, the hue represents the energy required to ionize the relevant ion (according to the color scale on the right). To increase the contrast between different systems, stronger systems are plotted in darker shades in the background and weaker systems are plotted in lighter shades in the foreground. The horizontal lines show the redshift range over which each ion can be detected. The grey shaded bands indicate areas where one or more transitions of the relevant ion fall in noisy spectral regions (see Section \ref{subsec:completeness}). The top panel shows a zoom-in on the redshift range \mbox{5.77 $\lesssim z \lesssim$ 6}, and illustrates the range of different kinematics and ionization structures found within the absorber sample. In this panel, every second component is plotted using a slightly darker shade for clarity. An animated version of this plot can be found in the \href{https://github.com/XQR-30/Metal-catalogue}{GitHub repository}.} \label{fig:redshift_space_spectra} 
\end{figure*}

\section{Sample Properties}\label{sec:absorber_properties}
\subsection{Overview}
The final catalog contains a total of 778 systems including \numCIV\ \CIV\ absorbers, \numMgII\ \MgII\ absorbers, and \numCII\ \CII\ absorbers (see Table \ref{linetable}). This sample represents a significant increase in the number of $z > 5$ absorbers observed at high spectral resolution and S/N compared to existing samples, as discussed in Section \ref{subsec:abs_path}. Figure \ref{fig:redshift_space_spectra} illustrates the significant statistical power of the full dataset. The bottom panel displays all absorption systems detected in each of the primary ions (\MgII, \FeII, \CIV, \SiIV, \CII\ and \NV) in all \mbox{E-XQR-30} spectra combined. The horizontal lines show the redshift range over which each ion can be detected, and the grey bands indicate regions where one or more transitions of the relevant ion fall in noisy spectral regions (see Section \ref{subsec:completeness}). Each system is plotted using a hue that represents the energy required to ionize the relevant ion (according to the color scale on the right). The number of systems is so large that many of them fall on top of one another in this collapsed redshift space. To ensure that all systems are visible, stronger systems are plotted in darker shades in the background with weaker systems in lighter shades in the foreground. 

\begin{figure*}
\centering
 \includegraphics[scale=1.1, clip = True, trim = 0 39 0 57]{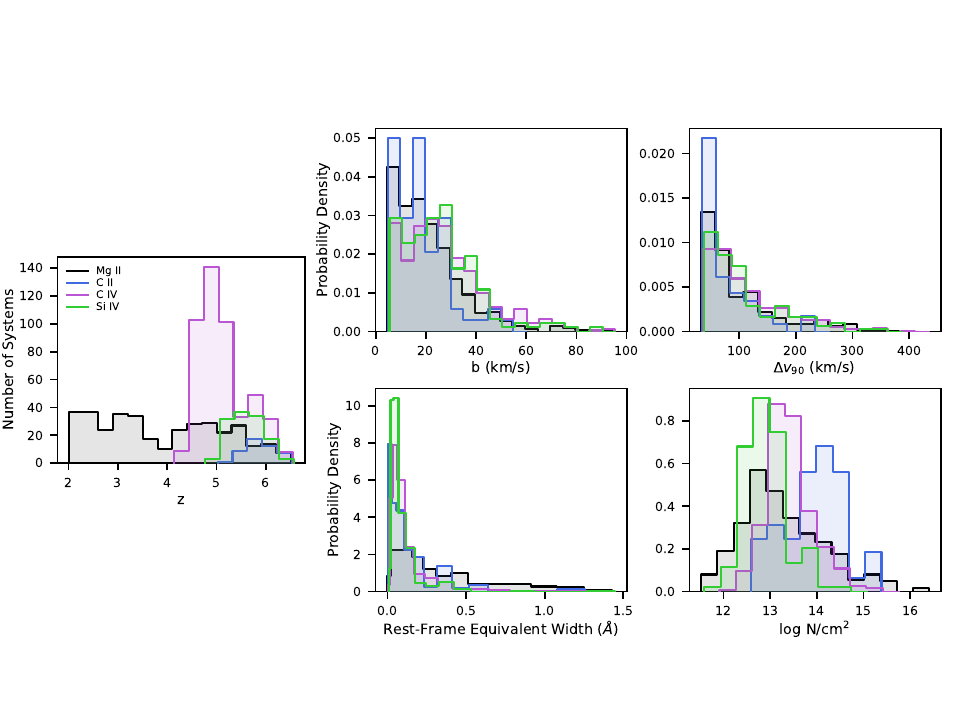}
 \caption{Distributions of redshift ($z$; left), $b$ parameter (upper middle), $\Delta v_{90}$ (upper right), rest-frame equivalent width ($W$; lower middle), and column density ($\log N$; lower right) for \MgII\ (black), \CII\ (blue), \CIV\ (purple), and \SiIV\ (green) absorbers, including both intervening and proximate systems. The equivalent width and $\Delta v_{90}$ are measured for the strongest line of each doublet. The  $z$, $\Delta v_{90}$, $W$ and $\log N$ distributions reflect the properties of absorption systems (described in Section \ref{subsec:system_grouping}) whilst the $b$ parameter distribution is necessarily based on measurements for individual Voigt profile components. The same bin boundaries are used for all ions but the distributions are artificially offset from one another for clarity.} \label{fig:absorber_stats} 
\end{figure*}

The top panel is a zoom-in on the region around \mbox{5.77 $\lesssim z \lesssim$ 6} and illustrates the kinematics and ionization structure of a few example systems. In this panel, every second component is plotted using a slightly darker shade to better highlight the locations of individual components. The absorption systems show a wide variety of properties, including narrow single component low-ionization systems seen only in \MgII, \FeII\ and \CII, multi-component high-ionization systems seen in \CIV, \SiIV, and \NV, and multi-phase systems detected in both low and high-ionization transitions. The rich multi-phase nature of this dataset will provide crucial insights into the ionization state of the CGM during the later stages of reionization. We observe absorption lines with a variety of strengths, from strongly saturated systems spanning a wide velocity range to very weak systems just above the detection limit. The large equivalent width range will enable us to robustly measure the properties of systems down to low metallicities and/or column densities.

\subsection{Absorber Statistics}\label{subsec:absorber_stats}
\subsubsection{Primary Ions}
The physical properties of the \MgII, \CIV, \SiIV\ and \CII\ absorbers are summarized in Figure \ref{fig:absorber_stats}. The redshift, equivalent width, $\Delta v_{90}$ and $\log N$ distributions show the properties of absorption systems, whilst the $b$ parameter distribution necessarily reflects the properties of individual Voigt profile components.

The left-most panel shows the distributions of absorber redshifts. \MgII\ absorbers (black) are detected relatively uniformly across the probed redshift range, with a couple of dips corresponding to wavelength regions with elevated contamination from skyline and/or telluric residuals (see Section \ref{subsec:completeness} and Figure \ref{fig:completeness_wavelength}). This dataset therefore allows us to self-consistently probe the evolution of \MgII\ absorption over 2.5~Gyr of early cosmic history. \SiIV\ (green) and \CII\ (blue) absorbers do not show any strong variations in incidence as a function of redshift, which may be due to the small redshift intervals over which these ions are accessible. The incidence of \CIV\ absorbers (purple) peaks strongly at $z\sim$~4.9 before declining rapidly towards higher redshifts. This is caused by a combination of the difference in resolution and S/N between the XSHOOTER VIS and NIR spectra, and an intrinsic drop in the occurence of \CIV\ absorbers at $z\gtrsim$~5 (e.g. \citealt{Songaila01, Simcoe06, RyanWeber06, RyanWeber09, DOdorico13, Diaz16, Codoreanu18, Meyer19, DOdorico22}; see Davies et al. submitted for a detailed analysis of \CIV\ absorber statistics using the E-XQR-30 sample). The decline in \CIV\ absorption at lower redshifts is the result of \CIV\ shifting into the inaccessible \Lya\ forest region.

The $b$ parameter distributions for all four ions (upper middle panel) show broad peaks at \mbox{10~--~20~\kms} with prominent tails towards larger values. The fraction of systems in the tail (with \mbox{$b \gtrsim$~30~\kms}) is higher for \CIV\ and \SiIV\ \mbox{(32~--~35\%)} than \MgII\ and \CII\ \mbox{(12~--~21\%)}, consistent with the expectation that higher ionization absorbers should arise from warmer and more turbulent material on average \citep[see also][]{Wolfe93, Rauch96, Lehner14, Muzahid14, Fox15, Cooper19, Pradeep20}. We do not observe strong peaks in the lowest $b$ parameter bin because there were relatively few cases where the $b$ parameter could not be reliably measured and had to be fixed to the minimum measurable value. In other words, the majority of components are at least partially resolved in our data.

The $\Delta v_{90}$ distributions (upper right panel) peak at \mbox{35~--~60~\kms}, indicating that the most common type of absorption system is a single Voigt component with a $b$ parameter of \mbox{5~--~20~\kms}. Systems with multiple kinematic components are also prevalent as indicated by the long tails which extend up to velocity widths of 400~\kms. The $\Delta v_{90}$ distributions are similar for all ions, with a slight preference towards lower velocity widths in \CII\ (and to a lesser extent \MgII) absorbers which may reflect their typically lower $b$ parameters.

The rest-frame equivalent width distributions (lower middle) similarly peak at \mbox{$W \simeq$~0.1\AA} with long tails towards larger values. The \SiIV\ distribution is most strongly concentrated in the weak absorber ($W <$~0.2\AA) regime. On the other hand, \MgII\ shows the flattest equivalent width distribution with a prominent tail extending to the ultra-strong regime ($W >$~1\AA). Strong \MgII\ absorbers trace the evolution of the cosmic star formation rate density \citep[e.g.][]{Steidel92, Matejek12, Zhu13, Chen17, Mathes17, Zou21} and are often associated with star-forming galaxies \citep[e.g.][]{Zibetti07, Noterdaeme10, Menard11}. They have been shown to trace extended galaxy disks \citep[e.g.][]{Zabl19}, outflows \citep[e.g.][]{Nestor11, Nielsen13, Rubin14, Lan18, Schroetter19}, and galaxy group environments \citep[e.g.][]{Rubin10, Gauthier13, Nielsen18, Dutta20, Lee21, Guha22}. We convert $W$ to N(H~\textsc{i}) using the empirical scaling relation published in \citet{Lan17} and find that 56/\numMgII of the \MgII\ absorbers could potentially be associated with Dampled Lyman Alpha (DLA) absorption. However, the scaling relation is constrained using measurements over \mbox{0 $\lesssim z \lesssim$~4.5} and therefore the extrapolation to higher redshifts is uncertain.

The high equivalent width \MgII\ absorbers do not dominate the high end of the $\log N$ distribution (lower right) due to the large oscillator strength of \MgII. Instead, the majority of systems with \mbox{$\log N >$ 14} are \CII\ absorbers. The \CIV\ systems generally have column densities intermediate between those of \CII\ and \MgII, and the \SiIV\ absorbers have the lowest median column density.

\begin{figure*}
\centering
 \includegraphics[scale=1, clip = True, trim = 0 0 0 0]{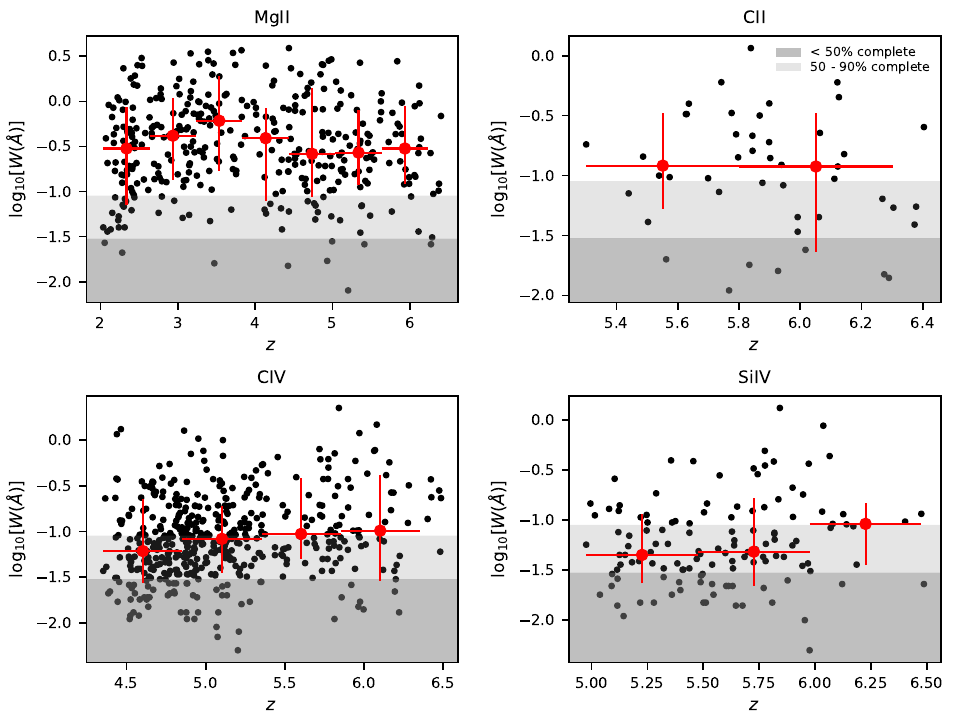}
 \caption{Rest-frame equivalent width as a function of redshift for \MgII\ (top left), \CII\ (top right), \CIV\ (bottom left) and \SiIV\ (bottom right) absorbers. Black markers show measurements for individual systems. Red markers and error bars indicate the median and 1$\sigma$ spread of values in redshift bins. The dark grey and light grey shaded areas indicate the regions where the sample is $<$~50\% and 50~--~90\% complete, respectively.} \label{fig:ew_z} 
\end{figure*}

Figure \ref{fig:ew_z} shows plots of $W$ as a function of $z$ for all four ions. Black markers show measurements for individual systems, and red points and error bars show the median and 1$\sigma$ spread in bins of $z$. The dark and light grey colored regions indicate areas where the sample is $<$~50\% and 50~--~90\% complete (see Section \ref{subsec:completeness} for details). There is a large scatter in the rest-frame equivalent width at fixed redshift for all ions, and there is no evidence for any significant variation in the typical $W$ probed at different redshifts. 

\subsubsection{Associated Transitions}
We detect a significant number of additional transitions associated with the primary absorption systems. In descending order, we find 80 systems with \AlII\ absorption, 77 with \SiII, 46 with \MgI, 29 with \OI\ (all with associated \CII), 16 with \AlIII, 8 with \MnII, 6 with \CII*, 4 with \CaII, 4 with \SII, 3 with \CrII, 2 with \SiII*, 2 with Ti~\textsc{II}, and 2 with \ZnII. 

\subsection{Total Absorption Path}\label{subsec:abs_path}
A key quantity for absorption line surveys is the absorption path $\Delta X$, which gives the redshift search interval in a comoving coordinate system. The absorption distance between \mbox{$z$~=~0} and \mbox{$z$~=~$z$} is given by:
\begin{equation}\label{eqn:absorption_path}
 X(z) = \frac{2}{3 \Omega_m} \left(\Omega_m \left(1 + z\right)^3 + \Omega_\Lambda \right)^{1/2}
\end{equation}
\citep[e.g.][]{Bahcall69}. Throughout this paper, all mentions of absorption path refer to the total $\Delta X$.

The total absorption path for a given ion in a spectrum with no masked regions can be computed by simply taking the difference between the absorption distances at the maximum and minimum probed redshifts. Table \ref{linetable} lists the total absorption path for each of the six primary ions, for the full E-XQR-30 sample as well as the XQR-30 quasars alone. We also list the path lengths for intervening absorbers which are relevant for studies of absorber statistics. In all cases, \textit{regions that were masked due to bad skyline or telluric contamination were subtracted from the absorption path} (see Section \ref{subsec:system_search}). Regions dominated by BAL features are also excluded when measuring $\Delta X$ because the completeness of narrow absorbers in these regions is poorly defined, making them difficult to incorporate in statistical measurements. The BAL regions were identified by searching for contiguous troughs with normalized flux less than 0.9 over a velocity interval of at least 2000~\kms\ (\citealt{Bischetti22} and submitted) and are listed in Table \ref{bal_region_table}. 

The sample presented in this paper represents a significant increase in the probed absorption path for \CIV\ and \SiIV\ absorbers at \mbox{$z \gtrsim$ 5}, which are key probes of the ionizing photon background near the end of the epoch of reionization \citep[e.g.][]{Songaila95, Songaila98, Boksenberg03, Bolton11, Finlator16, Doughty18, Graziani19}. Prior to this work, the most extensive search for \SiIV\ at $z \gtrsim$~4.9 covered an absorption search path of \mbox{$\Delta X$~=~32.7} \citep{DOdorico22}, which is a factor of $\sim$~4 smaller than the intervening path length covered by the current sample. For \CIV, our search path for intervening absorbers is a factor of $\sim$~2.6 ($\sim$~1.4) larger than any previous survey at $z\gtrsim$~5.2 \mbox{(4.3 $\lesssim z \lesssim$ 5.2)} (see Davies et al. submitted). 

The current sample also significantly increases the number of high-redshift, low-ionization absorbers measured at high S/N and spectral resolution. The largest survey to date of \MgII\ absorption at high redshift contains 38 quasars at $z\gtrsim$~5.8 observed at a spectral resolution of $\sim$~50~\kms\ \citep{Chen17}. Similarly, the largest survey of \CII\ absorption at high redshift comprises 47 quasars at $z\gtrsim$~5.7 observed at \mbox{$\sim$~50~\kms} resolution with a S/N of \mbox{$\sim$~5~--~10} \citep{Cooper19}. The largest sample of \OI\ and \CII\ absorbers observed at $\sim$~30~\kms\ resolution contains 39 quasars at $z\gtrsim$~5.8 \citep{Becker19}. 21/39 of these quasars are part of our sample, and adding the rest of the quasars in our sample increases the absorption path for $z\sim$~6 \OI\ and \CII\ absorbers by $\sim$~50\%. 

\section{Completeness And False-Positive Rate}\label{sec:completeness}
In order to recover the intrinsic statistical properties of absorber populations it is necessary to quantify the completeness (the fraction of real absorption systems that are recovered) and false-positive rate (the fraction of recovered systems that are spurious) in the sample. These quantities depend on both the data quality and analysis method and are impacted to different degrees by the automated system finding and filtering, visual inspection, and Voigt profile fitting. We produce a set of mock spectra for which the absorber properties are known a-priori (described in Section \ref{subsec:mock_spectra}), and process these spectra using the same steps applied to the observed data (outlined in Section \ref{sec:method}) to characterize the completeness as a function of equivalent width, redshift, column density and $b$ parameter for different ions (Section \ref{subsec:completeness}), as well as the overall false-positive rate in the sample (Section \ref{subsec:false_positive}).

\subsection{Mock Spectra}\label{subsec:mock_spectra}
Each mock spectrum is constructed by taking the continuum fit for one of the observed spectra, perturbing it by the measured error spectrum, and then inserting \MgII, \FeII, \CII, \CIV, \SiIV, and \NV\ absorption systems with a range of redshifts, column densities, and $b$ parameters. For \CII\ systems we also insert \OI, \SiII, \AlII, \MgII\ and \FeII\ lines, with column density ratios determined by fitting scaling relations to existing measurements of low-ionization $z>$~5 absorbers from \citet{Cooper19}. 

This method does not account for systematic errors originating from skyline or telluric subtraction. However, the wavelength regions with the worst systematics in the observed spectra are masked when searching for absorption systems (see Section \ref{subsec:system_search}) and are not included in the survey absorption path (see Section \ref{subsec:abs_path}), so we do not characterize the completeness in these regions. This approach also assumes that the error on the continuum fit is negligible, which is a reasonable assumption for non-BAL quasars but breaks down in wavelength regions with prominent BAL features. Therefore, the completeness statistics presented in this section only apply to absorbers falling in wavelength regions that are not masked and are not coincident with BAL features.

The number of systems inserted for each ion is set so that the automatic line finder identifies approximately the same number of candidate absorption systems in the mock spectra as in the observed spectra. This is important because the number of absorption lines per unit wavelength directly influences the probability of blending and the chance alignment of unassociated transitions. The inserted systems are uniformly distributed in redshift, from the minimum redshift where each ion is detectable to 5000~\kms\ above the quasar redshift.

The $b$ parameter of each inserted system is randomly selected from three possible values that approximately represent the 16th, 50th and 84th percentile of the values measured from the Voigt profile fitting. For the high-ionization transitions these values are 13~\kms, 24~\kms, and 47~\kms, while for the low-ionization transitions they are 7~\kms, 18~\kms\ and 32~\kms. The $\log N$ values are distributed uniformly over intervals chosen to span the full completeness range from 0\% to 100\% for each ion. 

To ensure robust statistics we produce 20 mock spectra per quasar which all have the same noise statistics but differing populations of inserted absorbers (resulting in a total of 840 mock spectra). All 840 mock spectra were run through the automatic line finding and filtering procedures described in Sections \ref{subsec:system_search} and \ref{subsec:auto_filtering} to produce preliminary catalogs and calculate the completeness and false-positive rate as a function of absorber properties. Visual inspection and Voigt profile fitting were performed on 1/20 spectra for each quasar to characterize the impact of these steps on the final completeness and false-positive rate (see Section \ref{subsec:false_positive}). 

We consider an inserted system to be successfully recovered if the mock catalog contains a system of the same ion within 50~\kms\ of the inserted redshift. The 50~\kms\ window was chosen by manually matching inserted and recovered systems for a single mock spectrum and calculating the velocity offsets between the inserted and recovered redshifts. The 50~\kms\ window is large enough to correctly match the majority of recovered systems to the corresponding inserted systems, but sufficiently small to avoid spurious matches. The largest velocity offsets occur when multiple strongly blended absorption components are recovered as a single system, with a redshift that may be significantly different from the central redshifts of the individual absorption components.

\begin{figure*}
\centering
 \includegraphics[scale=1]{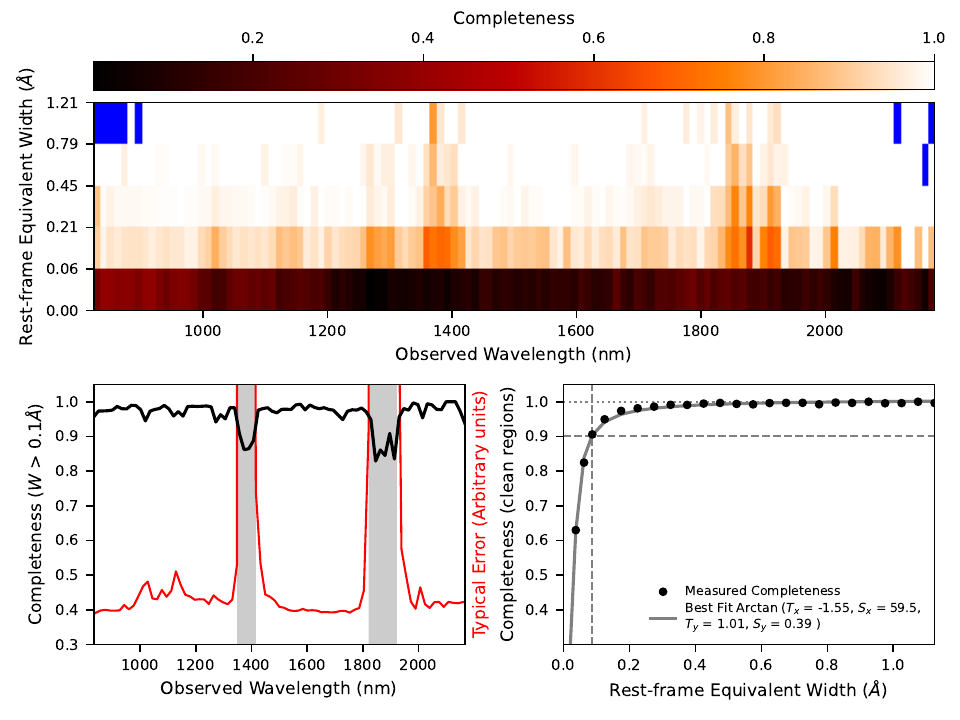}
 \caption{Top: Completeness as a function of rest-frame equivalent width ($W$) and observed wavelength for all ions combined. We emphasize that these statistics are computed excluding the masked spectral regions listed in Table \ref{masked_region_table}. The spectral range is divided into 120 bins each with a width of $\sim$~11\AA. The equivalent width bins are defined such that the square roots of the bin boundaries are equally spaced. Bins that have fewer than three systems are shown in blue. The completeness is a strong function of equivalent width and also dips at around 1400nm and 1900nm in the gaps between the $J$, $H$ and $K$ observing bands where sky noise is elevated. Bottom left: We measure the extents of the contaminated wavelength bands by plotting the completeness (black) and typical flux error (red) as a function of wavelength for systems with $W >$~0.1\AA. These systems are chosen because their overall completeness is high, meaning that any dips in completeness can be robustly identified. The low completeness regions are shown as grey bands and cover \mbox{1348~--~1416~nm} and \mbox{1821~--~1922~nm}. Bottom right: Completeness as a function of $W$ for clean wavelength regions (excluding the grey bands in the bottom left panel). The dependence is fit using an arctan function (Equation \ref{arctan_eqn}) with best-fit parameters given in the legend. The completeness exceeds 90\% (50\%) for $W \gtrsim$~0.09~\AA\ (0.03\AA).} \label{fig:completeness_wavelength} 
\end{figure*}

\subsection{Completeness Statistics}\label{subsec:completeness}
The sample completeness for a given ion varies most strongly with equivalent width and column density but is also dependent on the $b$ parameter (because a narrower line will be detected at higher peak S/N than a broader line at fixed $\log N$) and the redshift (because the noise level is strongly wavelength-dependent due to the presence of significant skyline contamination in some wavelength regions). In the following sections we explore how the completeness depends on each of these parameters.

\subsubsection{Completeness as a Function of Equivalent Width and Redshift}
Figure \ref{fig:completeness_wavelength} shows how the completeness varies as a function of rest-frame equivalent width and observed wavelength for all ions combined. The color scale reflects the completeness in each region of 2D parameter space, with blue patches indicating regions that do not contain sufficient data to make meaningful measurements. The completeness is generally very high ($\gtrsim$~90\%) for systems with $W \gtrsim$~0.2\AA, with the exception of two wavelength regions around $\sim$~1400~nm and $\sim$~1900~nm which correspond to regions of elevated sky contamination between the $J$, $H$ and $K$ observing bands. We re-iterate that the completeness in these regions is measured using only about half of the E-XQR-30 quasars because the remainder show strong skyline/telluric residuals which were masked (see Table \ref{masked_region_table}). \textit{If we had not masked the most contaminated wavelength ranges, the completeness in the regions between the observing bands would be significantly lower (even for the relatively strong absorbers considered here) while the total absorption search path would be correspondingly larger}. The completeness also decreases at lower equivalent widths. 

To more accurately measure the extents of the wavelength regions that are most strongly impacted by low completeness, we compute the completeness as a function of wavelength using all absorbers with \mbox{$W >$~0.1\AA}, as shown in the bottom-left panel of Figure \ref{fig:completeness_wavelength}. We consider absorbers in this equivalent width range because their overall completeness is high, meaning that any dips in completeness due to elevated errors can be robustly quantified. We identified regions of low completeness by selecting wavelength channels where the average completeness is less than 90\%, and then expanding these regions to include any neighbouring channels with completeness below 95\%. The low-completeness regions, shown as grey bands, cover \mbox{1348~--~1416nm} and \mbox{1821~--~1922nm}.

The presence of low-completeness wavelength bands translates directly to reductions in the completeness of individual ions over the corresponding redshift intervals. The contaminated bands lie sufficiently far into the near-infrared that none of \CII, \CIV\ or \SiIV\ are shifted to those wavelengths within the redshift range probed by our sample. However, the low-completeness bands translate to dips in the \MgII\ completeness of our dataset over \mbox{3.81 $\leq z \leq$ 4.05} and \mbox{5.50 $\leq z \leq$ 5.86}. When quantifying the completeness statistics for \MgII\ in the following section we exclude these contaminated regions of redshift space.

Finally, we examine the completeness as a function of equivalent width (excluding the contaminated wavelength bands), as shown in the bottom right panel of Figure \ref{fig:completeness_wavelength}. The completeness rises steeply at $W \lesssim$~0.1\AA\ before plateauing at $W \gtrsim$~0.2\AA. The variation in completeness as a function of wavelength is well fit by an arctan function, parametrized as follows:
\begin{equation}\label{arctan_eqn}
{\rm Completeness}(W) = S_y \left(\arctan \left(S_x W + T_x \right) + T_y \right) 
\end{equation}
where $T_{x,y}$ and $S_{x,y}$ represent the translation and scaling of each axis, respectively. The best-fit is shown in grey and the parameters are listed in the figure legend. Based on this fit, the completeness exceeds 90\% (50\%) for $W \gtrsim$~0.09\AA\ (0.03\AA). 

\subsubsection{Completeness as a Function of Column Density and $b$ Parameter}
Next, we compute the completeness as a function of column density for \MgII, \CII, \CIV\ and \SiIV. The average error on the observed flux varies as a function of wavelength, so we calculate the completeness in redshift bins. For \CII, \CIV\ and \SiIV\ we use two redshift bins divided at the path-length-weighted mean redshift of the survey. For \MgII, we exclude the redshift ranges \mbox{3.81 $\leq z \leq$ 4.05} and \mbox{5.50 $\leq z \leq$ 5.86} where sky contamination was found to be most significant. 

\begin{figure*}
\centering
 \includegraphics[scale=1.0, clip = True, trim = 0 95 5 0]{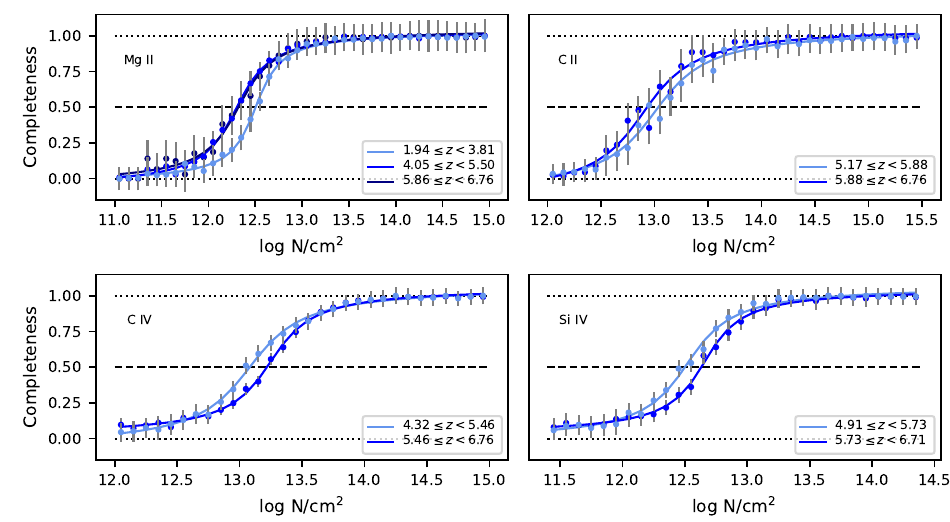}
 \caption{Completeness as a function of column density in different redshift bins for \MgII, \CII, \CIV\ and \SiIV\ absorbers. For \MgII, the completeness is computed in 3 redshift bins and we do not report values for the redshift ranges impacted by strong sky contamination (\mbox{3.81 $\leq z \leq$ 4.05} and \mbox{5.50 $\leq z \leq$ 5.86}). For \CII, \CIV\ and \SiIV\ the completeness is computed in two redshift bins divided at the path-length-weighted mean redshift of the survey. We quantify the variation in completeness as a function of $\log N$ by fitting arctan functions which are over-plotted in the same color as the completeness measurements for each redshift bin. The parameters of the arctan curves are listed in Table \ref{completeness_table}. } \label{fig:completeness_logN_z} 
\end{figure*}

The results are shown in Figure \ref{fig:completeness_logN_z}. In all cases, the completeness asymptotes towards 0 at low $\log N$ and 1 at high $\log N$, with an almost linear increase in the transition region. Therefore, we again characterize the completeness variations by fitting arctan functions which are plotted in the same color as the completeness measurements for each redshift bin. The best-fit parameters for the arctan functions are listed in Table \ref{completeness_table}. 
 
\begin{table}
\begin{nscenter}
\begin{tabular}{l|c|c|c|c|c}
\hline
 & &  &  &  & 50\%  \\ 
 & &  &  &  & Completeness  \\ 
 & &  &  &  & Limit  \\ 
subset & $T_x$ & $S_x$ & $T_y$ & $S_y$ & $\left(\log_{10}\left(\frac{\rm N}{\rm cm^{-2}}\right)\right)$ \\ \hline
\multicolumn{6}{c}{Mg II} \\ \hline
1.94 $\leq z <$ 3.81 & -62.8 & 5.02 & 1.48 & 0.34 & 12.50 \\ \hline
4.05 $\leq z <$ 5.50 & -45.4 & 3.70 & 1.38 & 0.36 & 12.30 \\ \hline
5.86 $\leq z <$ 6.76 & -43.7 & 3.55 & 1.44 & 0.35 & 12.32 \\ \hline
$b$ = 7\kms\ & -35.7 & 2.91 & 1.43 & 0.35 & 12.27 \\ \hline
$b$ = 18\kms\ & -77.4 & 6.34 & 1.54 & 0.33 & 12.21 \\ \hline
$b$ = 32\kms\ & -58.4 & 4.69 & 1.54 & 0.34 & 12.45 \\ \hline
\multicolumn{6}{c}{C II} \\ \hline
5.17 $\leq z <$ 5.88 & -35.5 & 2.73 & 1.26 & 0.37 & 13.04 \\ \hline
5.88 $\leq z <$ 6.76 & -35.5 & 2.75 & 1.19 & 0.39 & 12.93 \\ \hline
$b$ = 7\kms\ & -33.5 & 2.62 & 1.06 & 0.40 & 12.84 \\ \hline
$b$ = 18\kms\ & -34.7 & 2.69 & 1.12 & 0.40 & 12.90 \\ \hline
$b$ = 32\kms\ & -51.5 & 3.95 & 1.45 & 0.34 & 13.05 \\ \hline
\multicolumn{6}{c}{C IV} \\ \hline
4.32 $\leq z <$ 5.46 & -40.9 & 3.13 & 1.37 & 0.36 & 13.08 \\ \hline
5.46 $\leq z <$ 6.76 & -52.0 & 3.93 & 1.61 & 0.33 & 13.22 \\ \hline
$b$ = 13\kms\ & -53.4 & 4.06 & 1.61 & 0.33 & 13.14 \\ \hline
$b$ = 24\kms\ & -52.6 & 3.98 & 1.58 & 0.34 & 13.17 \\ \hline
$b$ = 47\kms\ & -60.0 & 4.49 & 1.66 & 0.33 & 13.34 \\ \hline
\multicolumn{6}{c}{Si IV} \\ \hline
4.91 $\leq z <$ 5.73 & -46.4 & 3.70 & 1.50 & 0.35 & 12.50 \\ \hline
5.73 $\leq z <$ 6.71 & -56.7 & 4.48 & 1.64 & 0.33 & 12.63 \\ \hline
$b$ = 13\kms\ & -67.8 & 5.39 & 1.72 & 0.32 & 12.56 \\ \hline
$b$ = 24\kms\ & -53.4 & 4.24 & 1.53 & 0.34 & 12.57 \\ \hline
$b$ = 47\kms\ & -65.3 & 5.10 & 1.72 & 0.32 & 12.77 \\ \hline
\end{tabular}
\caption{Parameters of the arctan functions (Equation \ref{arctan_eqn}) fit to the completeness as a function of $\log N$ for \MgII, \CII, \CIV\ and \SiIV. We fit the completeness in bins of redshift and for each of the individual $b$ parameter values that were selected from when inserting mock systems (see Section \ref{subsec:mock_spectra}).}\label{completeness_table}
\end{nscenter}
\end{table}
 
We find that the sub-sample of \MgII\ absorbers at $z\geq$~4.05 is slightly more complete at intermediate $\log N$ than the subsample at $z\leq$~3.81. This may be because the lower redshift systems fall at shorter observed wavelengths where they are more likely to be blended with other absorption lines, making them more difficult to recover. For \CII\ we do not find any significant difference in completeness between the two redshift bins. For \CIV\ and \SiIV\ the higher redshift bins are less complete at intermediate $\log N$, likely due to the increase in the average flux error towards longer wavelengths. 

\begin{figure*}
\centering
 \includegraphics[scale=1.0, clip = True, trim = 0 95 5 0]{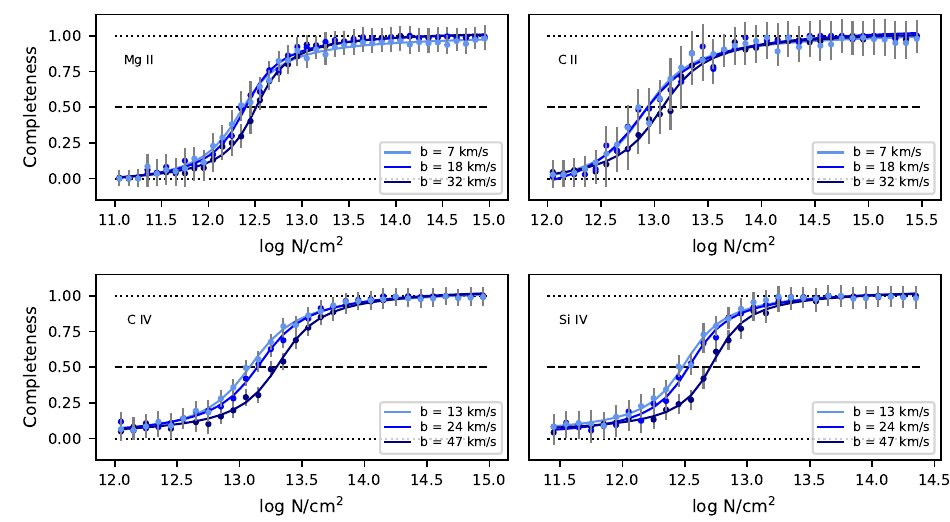}
 \caption{Same as Figure \ref{fig:completeness_logN_z} but separated by $b$ parameter. For each ion, the three $b$ parameter values are those that were selected from when inserting absorption systems into the mock spectra (see Section \ref{subsec:mock_spectra}).} \label{fig:completeness_b} 
\end{figure*}

The inserted systems for each ion have one of three $b$ parameter values, so we can also fit the completeness as a function of $\log N$ for each $b$ parameter as shown in Figure \ref{fig:completeness_b}. As before, the best-fit parameters are listed in Table \ref{completeness_table}. For all ions we find that completeness at intermediate $\log N$ decreases with increasing $b$ parameter, which is to be expected because the S/N per spectral channel is lower when the absorption is spread over a larger range of velocities. 

We note that completeness statistics for ions not included in Table \ref{completeness_table} can be computed using the equivalent width scaling given in the bottom right panel of Figure \ref{fig:completeness_wavelength}.

\subsection{False-Positive Rate and Impact of Visual Inspection}\label{subsec:false_positive}
\begin{figure}
\centering
 \includegraphics[scale=0.55, clip = True, trim = 10 0 0 0]{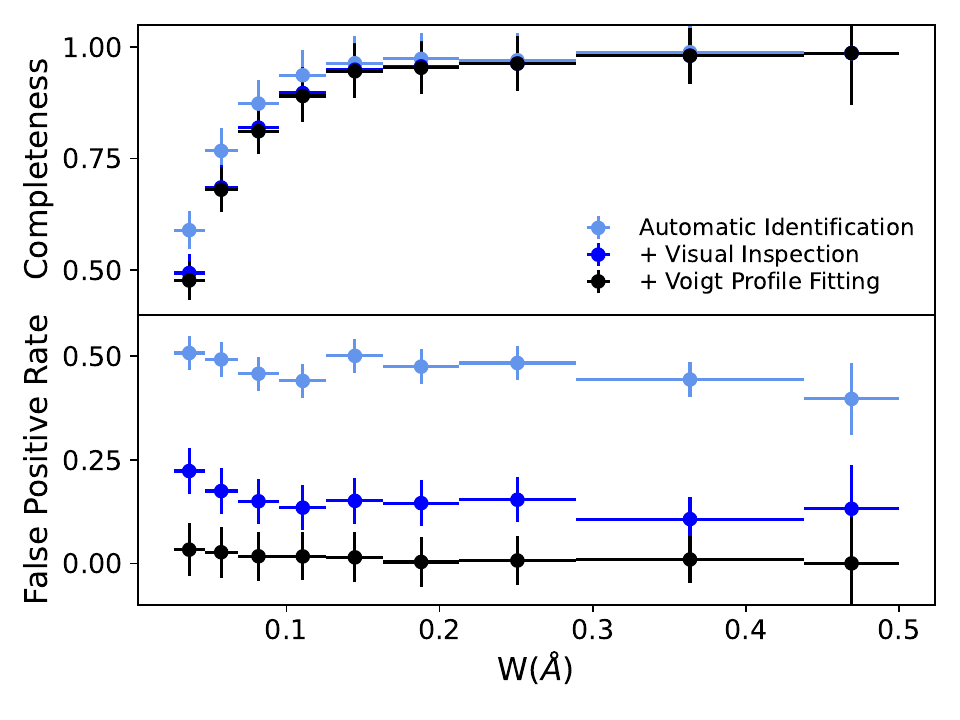}
 \caption{Completeness and false positive rate as a function of rest-frame equivalent width at different stages of the data processing: after automatic system identification and filtering (light blue), after visual inspection (blue), and after Voigt profile fitting (black). The visual inspection and fitting remove $<$~10\% of real systems but $\gtrsim$~90\% of spurious candidates, reducing the false positive rate from $\sim$~40~--~50\% to $<$~5\% in the final catalog.} \label{fig:checking_impact} 
\end{figure}

The completeness statistics presented in the previous section were computed using the automatically produced preliminary catalogs for the full set of 840 mock spectra. However, it is possible that the visual inspection and Voigt profile fitting applied to the observed spectra could modify the completeness. To investigate this, we use the subset of 42 mock spectra that were fully processed to calculate the completeness as a function of equivalent width at three different stages of the data processing: after the automatic processing, after the visual inspection, and after the Voigt profile fitting. The results are shown in the top panel of Figure \ref{fig:checking_impact}. For equivalent widths below $\sim$~0.1\AA\ the visual inspection reduces the completeness by $\lesssim$~10\%, while for stronger systems the impact is negligible. The Voigt profile fitting does not significantly impact the completeness at any equivalent width.

In contrast, the bottom panel of Figure \ref{fig:checking_impact} shows that the visual inspection and Voigt profile fitting play crucial roles in reducing the false positive rate in the final sample. The visual inspection reduces the false-positive rate from \mbox{$\sim$~40~--~50\%} to $\sim$~20~--~30\% at all equivalent widths whilst removing very few real systems, therefore filtering the catalog very effectively. The remaining false positives originate from cases where a single absorption line has multiple associated candidates transitions. These are removed in the Voigt profile fitting phase, resulting in a final false positive rate of $<$~5\% at all equivalent widths. Contamination from false positives can therefore be treated as negligible in the final catalog.

\subsection{Cross-check with Previously Published Catalogs}\label{subsec:cross_check}
Many of the absorption systems reported in this paper are published here for the first time. However, absorption line catalogs have previously been published for the archival spectra as well as for some of the \mbox{XQR-30} quasars using lower quality spectra. We examine how the completeness of our sample compares to previous studies by cross-checking our catalog with absorber lists previously published for the same quasars.

\citet{Becker19} catalogued low-ionization absorbers in 21 of the quasars in our sample. They report 11 \OI\ absorbers, all of which appear in our catalog associated with automatically identified \CII\ systems. The improved spectra used in this analysis enable us to identify one additional \OI\ absorber in the set of overlapping spectra.

\citet{Cooper19} searched for \CII\ absorbers in 11 of the quasars in our sample. They report robust detections of \CII\ absorption in five systems, all of which are recovered by our automated line finding procedure and appear in the final catalog. We also obtain $>$~3$\sigma$ detections of \CII\ absorption in four systems for which \citet{Cooper19} reported upper limits, as well as 11 additional systems not reported in \citet{Cooper19}.

\citet{Chen17} catalogued \MgII\ absorbers in 19 of the quasars in our sample. They find 70 \MgII\ absorbers, of which we recover 66. One missed system coincides with a region of elevated noise around the \MgII~$\lambda$2796 line, while the other three systems were found to be better explained by transitions of other ions at different redshifts. Our comprehensive, simultaneous search for low and high-ionization systems allows us to more robustly identify the origin(s) of individual absorption lines. We identify 95 additional \MgII\ systems in the quasars surveyed by \citet{Chen17}.

Catalogs of \CIV\ and \SiIV\ absorption have been published for 12/42 of the E-XQR-30 quasars \citep{Codoreanu18, Meyer19, DOdorico13, DOdorico22}. Where multiple catalogs have been published for a given spectrum we compare with the most recent one. The existing catalogs contain a combined total of 148 \CIV\ absorbers and \SiIV\ absorbers, of which we recover 129 (87\%). Of the unrecovered systems, 10/19 (53\%) are not found in the automatic line finding stage. 8/10 of the unidentified doublets have column densities at least \mbox{0.15 dex} below the completeness limit of our dataset, while the remaining 2/10 lie in regions with significant skyline residuals. Most of these systems were previously identified by eye. 4/19 (21\%) of the unrecovered systems were found to be better explained by transitions of other ions at different redshifts. Finally, 3/19 (16\%) of the unrecovered systems were rejected by the automatic filtering algorithm due to blending with other much stronger absorption systems, and 2/19 (11\%) were rejected by the checkers due to low S/N. Systems that were missed in the initial catalog generation but were plausibly present at low S/N in our data were manually added, with catalog flags used to indicate that they are not part of the automatically identified primary sample (see Section \ref{subsec:flags}). Our catalog adds approximately 50 \CIV\ systems and 10 \SiIV\ systems to those previously reported for the same quasars.

\section{Description of Released Products}\label{sec:catalog_desc}
This paper is accompanied by a GitHub repository (\href{https://github.com/XQR-30/Metal-catalogue}{\url{https://github.com/XQR-30/Metal-catalogue}}) which contains the metal absorber catalogs and all of the data required to use these catalogs for scientific studies. Specifically, the repository includes:
\begin{itemize}
 \item A metal absorber catalog for each quasar, as well as a full catalog for the entire sample (provided in CSV format).
 \item Plots of the Voigt profile fits for individual absorption systems (provided in PDF format).
 \item Machine-readable versions of Tables \ref{redshift_table}, \ref{masked_region_table} and \ref{bal_region_table} (provided in CSV or JSON format).
 \item A \textsc{Python} tool to calculate the absorption path over a given redshift range for any transition or primary ion and any value of $\Omega_m$.
 \item Quasar continuum fits (provided in PDF format).
\end{itemize}

A condensed excerpt of the absorber catalog for \mbox{J0408-5632} is shown in Table \ref{table:catalog} to illustrate the contents of the data columns. The plots of the Voigt profile fits for the absorption systems listed in Table \ref{table:catalog} are shown in Figure \ref{fig:fitPlots}.

\subsection{Absorber Properties}\label{subsec:column_desc}
Columns 1-13 of Table \ref{table:catalog} give the IDs and physical properties of the detected absorption systems. In the full catalog, there is an additional entry at the start of each row (labelled Column 0) indicating which quasar the relevant absorption system is associated with. We illustrate the format of the table entries using System 9 in \mbox{J0408-5632} as an example. Figure \ref{fig:fitPlots} shows that this system is detected in \MgII, \FeII, \SiII, \AlII, and \CIV. The low-ionization species all share the same central two-component kinematic structure, although only the blue-most component is fit to \FeII\ due to the low S/N (the detection is listed as an upper limit). \MgII\ also shows a weaker redshifted component not seen in the other ions. \CIV\ is fit completely independently of the low-ionization species with four Voigt components. 

The first row of Table \ref{table:catalog} lists the overall properties of the \MgII\ absorption: the optical-depth-weighted mean absorption redshift, the rest-frame equivalent width of the strongest transition (in this case \MgII~$\lambda$2796), the total $\log N$, calculated by summing the column densities of the individual Voigt components, and $\Delta v_{90}$. The $b$ parameter is not listed because it can only be measured for individual components. The next row lists the equivalent width of \MgII~$\lambda$2803. If further \MgII\ transitions were detected, their equivalent widths would be listed in subsequent rows. Similar entries for \FeII, \SiII, \AlII\ and \CIV\ are found further down the table.

The next three rows in the table list the column density, $\Delta v_{90}$ and $b$ parameter of the individual Voigt profile components which are labelled 9a to 9c. In these IDs, the `9' denotes the system number and the letters are the component labels. These labels make it possible to link components with the same kinematics across different ions. If we examine the properties of component 9a for \MgII, \FeII, \SiII\ and \AlII, we see they all have the same $b$ parameter and redshift, indicating that their kinematics were tied during fitting. It was not always possible to fit all high-ionization and all low-ionization ions using the same kinematic structure (due to e.g. saturation, low S/N, or intrinsic differences between the velocity profiles; see Section \ref{subsec:voigt_fitting}). In these cases, the component labels (letters) would differ between ions. We note that \CIV, which was fit independently from the low-ionization species, has a different set of components labelled 9d to 9g.

If the optical depth measured from the weakest transition of a given ion is $\tau >$~2, then we classify the line as saturated and flag the column density measurement with `S' (saturated) in Column 10. If no transitions of a given ion are detected at $\geq$~3$\sigma$, then the reported equivalent width values are 3$\sigma$ upper limits. The $W$ upper limits are converted to $\log N$ values assuming that the absorption falls on the linear part of the curve of growth, and the minimum of the $\log N$ values for the individual transitions is reported as the overall upper limit for that ion. In such cases, the $W$ and $\log N$ values are flagged with `U' (upper limit) in Columns 7 and 10, respectively.

\subsection{Sample Flags}\label{subsec:flags}
The remaining three columns of the table provide various flags to assist users in determining which subset of absorbers should be used for a given science case. The completeness and false positive rate statistics presented in Sections \ref{subsec:completeness} and \ref{subsec:false_positive} only apply to systems that are recovered by the automatic line finder, pass the visual inspection, and do  not lie in masked wavelength regions or BAL regions. Objects that do not pass the visual inspection have a much higher likelihood of being false positives (see Figure \ref{fig:checking_impact}). The completeness in the masked wavelength regions is low due to significant sky contamination (see Figure \ref{fig:completeness_wavelength}). There is no straightforward way to quantify the impact of BAL absorption on our ability to recover underyling narrow absorption. For these reasons, Column 14 gives four flags indicating, from left to right, whether each ion in each absorption system 1) was recovered by the automatic line finder, 2) passed the visual inspection 3) does not lie in a masked wavelength region, and 4) does not lie in a BAL region. 

In addition, the completeness is only well defined for the `primary' ions that we search for in the initial candidate finding stage: \MgII, \FeII, \CII, \CIV, \SiIV, and \NV, as well as \OI\ and \SiII~$\lambda$1260 because they are required for the acceptance of \CII\ systems when they lie within the probed wavelength region (see Section \ref{subsec:system_search}). Absorption in other ions can only be recovered at redshifts where absorption is already detected in one or more of the primary ions, making the completeness poorly defined.

With this in mind we define a `primary' sample consisting only of primary ions in systems that meet all of the outlined criteria (all flags in Column 14 set to `Y'). The completeness and false positive rate for this sample are well understood, and only objects in this primary sample should be used for statistical studies requiring accurate completeness estimates. Column 15 indicates whether or not a given ion in a given system is part of the primary sample. The final column in the table indicates whether each absorption system is proximate (P) or intervening (I). 

\subsection{Example Usage: Redshift Evolution of Metal Ions}\label{subsec:example}
The metal absorber catalog presented in this work significantly expands on existing samples of metal absorbers at $z\gtrsim$~5 (see Section \ref{subsec:cross_check}) and is therefore well suited to study the evolution of metal absorber properties over \mbox{5 $\lesssim z \lesssim$ 6}. The evolution of \CIV\ absorber properties over \mbox{4.3 $\lesssim z \lesssim$ 6.3} is presented in an accompanying paper by Davies et al. (submitted), and the redshift evolution of low-ionization absorbers will be examined in a follow-up paper by Sebastian et al. (in prep).

To illustrate how the catalog can be used to quantify the evolution in metal absorbers properties, we outline the basic procedure to measure the cosmic mass density of intervening \CIV\ absorbers ($\Omega_{\rm C\, IV}$) over an arbitrary redshift interval \mbox{$z_1$~--~$z_2$}:
\begin{itemize}
\item Extract $\log N$ values (Column 8) for all rows that have:
\begin{itemize}
 \item A valid Component ID in Column 2 (not `-')
 \item $z$ in Column 3 within the redshift interval of interest; i.e. \mbox{$z_1 \leq z \leq z_2$}
 \item `CIV' listed under Ion/Species in Column 4
 \item `Y' listed under Primary Sample in Column 15
 \item `I' listed under Proximate or Intervening in Column 16
\end{itemize}
 \item Remove components for which $\log N$ is less than the 50\% completeness limit listed in Table \ref{completeness_table}.
 \item Group the remaining components into bins of $\log N$ with a width of \mbox{d$\log N$ = 0.1}.
 \item For each bin, calculate the total $\log N$ of all components and correct for completeness using Equation \ref{arctan_eqn} and the parameters listed in Table \ref{completeness_table}. If the redshift interval of interest is large enough that it spans multiple redshift bins for the completeness correction, this step can be performed for each redshift bin individually.
 \item Sum the total $\log N$ from all redshift and $\log N$ bins to obtain the overall total column density.
\end{itemize}

\begin{sloppypar}
The final step is to determine the total absorption path over the interval \mbox{$z_1$~--~$z_2$}. For each quasar, the unmasked absorption search path is given by \mbox{$\Delta X$ = $X(z_{\rm upper}) - X(z_{\rm lower})$}, where $X(z)$ is defined in Equation \ref{eqn:absorption_path}. $z_{\rm lower}$ is the larger of $z_1$ and the minimum redshift at which \CIV~$\lambda$1548 falls redward of the quasar \Lya\ line; i.e. \mbox{$z_{\rm lower}$ = max($z_1$, $\lambda_{\rm Ly\alpha}$(1 + $z_{\rm em}$)/$\lambda_{\rm CIV ~ 1548} -$ 1)}. $z_{\rm upper}$ is the smaller of $z_2$ and 10,000~\kms\ blueward of the quasar redshift (the maximum redshift for intervening absorbers); i.e. \mbox{$z_{\rm upper}$ = min$\left(z_2, \, (1 + z_{\rm em})\times \exp \left(\frac{-10,000~{\rm km \, s^{-1}}}{c}\right) - 1\right)$
}\footnote{This conversion between d$v$ and d$z$ was adopted for consistency with \textsc{astrocook}. It is derived as follows: d$v/c \simeq$ d$\lambda/\lambda$ = dln$\lambda$, which gives \mbox{d$z$ = (1+$z_{\rm em}$)d$\lambda \simeq$ (1+$z_{\rm em})\exp$(d$v/c$)}.}. Once the unmasked absorption path is calculated, any masked regions (listed in Table \ref{masked_region_table}) and BAL regions (listed in Table \ref{bal_region_table}) that fall within the relevant wavelength range must be subtracted to obtain the masked absorption path. Finally, the masked absorption paths of all quasars can be summed to obtain the total absorption path. The GitHub repository includes a \textsc{python} routine to calculate the absorption path over a given redshift range for any transition or primary ion and any value of $\Omega_m$.
\end{sloppypar}

\section{Summary}\label{sec:summary}
We have presented a catalog of 778 metal absorption systems found in high quality XSHOOTER spectra of 42 bright quasars at \mbox{5.8 $\lesssim z \lesssim$ 6.6}. Thirty of the quasars were observed as part of the VLT Large Program \mbox{XQR-30} (PI V. D'Odorico), and the remaining 12 have archival spectra with similar S/N ($\gtrsim$~10 per 10~\kms\ spectral pixel at 1285\AA) and spectral resolution ($\sim$~30~\kms).

The absorber catalog was constructed by performing an automated search for \MgII, \FeII, \CII, \CIV, \SiIV\ and \NV\ absorption, automatically filtering the list of candidates using custom routines, searching for additional transitions at the redshifts of the candidate systems, visually inspecting the candidates to remove spurious systems, and fitting Voigt profiles to the absorption lines. Many of these steps make use of routines from the \textsc{Python} library \textsc{Astrocook} \citep{Cupani20}.

We quantify the completeness and false-positive rate in the absorber sample by producing a set of 840 mock spectra for which the absorber properties are known a-priori, and processing them using the same steps applied to the observed data. The final false-positive rate is negligible ($<$~5\%) at all equivalent widths, and the sample is 50\%~(90\%) complete for \mbox{$W$(rest) $\gtrsim$~0.03\AA\ (0.09\AA)}. We provide functional forms that can be used to calculate the completeness as a function of rest-frame equivalent width, and as a function of $\log N$ for \MgII, \CII, \CIV\ and \SiIV, in different redshift intervals and for a range of $b$ parameters.

The final catalog includes \numCIV\ \CIV\ absorbers, \numMgII\ \MgII\ absorbers, and \numCII\ \CII\ absorbers. The sample covers a factor of $\sim$~3~--~4 larger absorption path than any previously published search for \CIV\ and \SiIV\ absorbers at $z\gtrsim$~5, and significantly expands existing samples of low-ionization absorbers observed at high S/N and spectral resolution. The number of high S/N spectra observed at comparable spectral resolution is unlikely to increase significantly in the near future because the space density of luminous quasars drops rapidly at \mbox{$z \gtrsim$ 6} \citep[e.g.][]{Wang19, Ren21}. Therefore, the catalog presented here is a crucial legacy resource for studies of metal absorber properties during the later stages of reionization. The catalog is also a key reference sample for evolutionary studies of enriched gas in galaxy halos. The redshifts of the absorption systems span \mbox{2 $\lesssim z \lesssim$ 6.5}, enabling the self-consistent comparison of absorber properties all the way from the end of reionization to the peak epoch of star formation. 

The redshift evolution of \CIV\ absorber properties is presented in an accompanying paper by Davies et al. (submitted), and the evolution of low-ionization absorber properties will be published in a follow-up paper by Sebastian et al. (in prep). To maximize the scientific impact of this dataset, we have publicly released the complete absorption line catalog along with all of the information required to use the data for statistical analysis. This includes machine-readable versions of the published tables as well as a \textsc{Python} tool to calculate the total absorption path for any ion/transition and redshift interval.

The unprecented statistical power of this dataset makes it possible to robustly measure the typical properties of metal absorbers near the end of reionization, opening up a myriad of possibilities for ground-breaking research. These data have the potential to facilitate valuable new insights on the formation redshifts of Population III stars, the strength and hardness of the UV background at $z\sim$~6, the properties of star-formation driven outflows in the early Universe, and the formation redshifts of the first galaxies. By propelling our understanding of absorber properties at the current redshift frontier, this dataset will create a firm basis for future studies of the even more distant Universe with JWST and 30m-class telescopes. 

\section*{Acknowledgements}
We thank the referee for their valuable suggestions which improved the clarity of this paper. RLD acknowledges the support of a Gruber Foundation Fellowship research support grant. SEIB and RAM acknowledge funding from the European Research Council (ERC) under the European Union's Horizon 2020 research and innovation programme (grant agreement no. 740246 "Cosmic Gas''). GDB and YZ were supported by the National Science Foundation through grant AST-1751404. ACE and FW acknowledge support by NASA through the NASA Hubble Fellowship grants $\#$HF2-51434 and HF2-51448.001-A, awarded by the Space Telescope Science Institute, which is operated by the Association of Universities for Research in Astronomy, Inc., for NASA, under contract NAS5-26555. EPF is supported by the international Gemini Observatory, a program of NSF’s NOIRLab, which is managed by the Association of Universities for Research in Astronomy (AURA) under a cooperative agreement with the National Science Foundation, on behalf of the Gemini partnership of Argentina, Brazil, Canada, Chile, the Republic of Korea, and the United States of America. This research was supported by the Australian Research Council Centre of Excellence for All Sky Astrophysics in 3 Dimensions (ASTRO 3D), through project number CE170100013. Based on observations collected at the European Organisation for Astronomical Research in the Southern Hemisphere under ESO Programme IDs 0100.A-0625, 0101.B-0272, 0102.A-0154, 0102.A-0478, 084.A-0360(A), 084.A-0390(A), 084.A-0550(A), 085.A-0299(A), 086.A-0162(A), 086.A-0574(A),087.A-0607(A), 088.A-0897(A), 091.C-0934(B),  096.A-0095(A), 096.A-0418(A), 097.B-1070(A), 098.B-0537, 098.B-0537(A), 1103.A-0817, 294.A-5031(B), 60.A-9024(A). This research made use of NASA's Astrophysics Data System as well as \textsc{Astrocook} \citep{Cupani20}, \textsc{Astropy} \citep{Astropy18}, \textsc{Matplotlib} \citep{Hunter07}, \textsc{Numpy} \citep{Harris20}, and \textsc{Scipy} \citep{Scipy20}.

\section*{Data Availability}
The metal absorber catalogs, data tables, quasar continuum fits and accompanying \textsc{python} scripts are released with this paper and can be downloaded from this GitHub repository: \href{https://github.com/XQR-30/Metal-catalogue}{\url{https://github.com/XQR-30/Metal-catalogue}}. The reduced spectra will be released with the XQR-30 survey paper (D'Odorico et al., in prep.).

\bibliographystyle{mnras}
\bibliography{../bibliography/mybib}

\appendix

\section{Velocity shifting}\label{appendix:velshift}
\begin{figure*}
\centering
 \includegraphics[scale=1, clip = True, trim = 0 155 0 0]{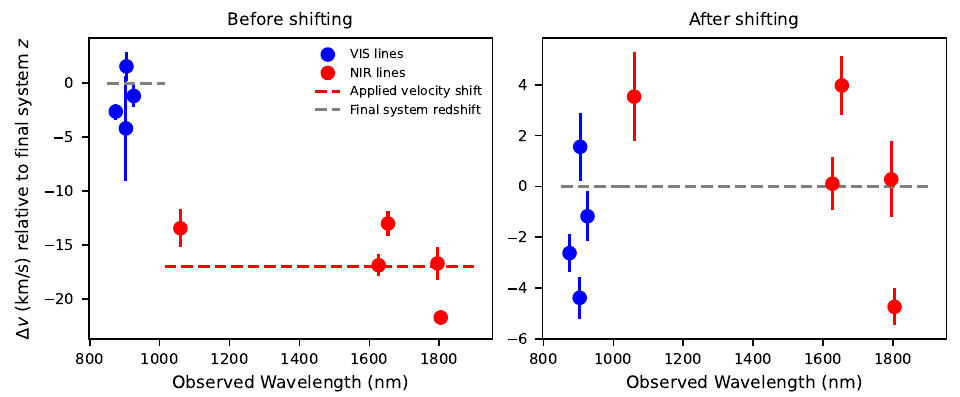}
 \caption{Left: Difference between the centroid velocities of VIS and NIR transitions (blue and red markers, respectively) as a function of observed wavelength for a single-component low-ionization absorption system at $z$ = 5.939 in the spectrum of SDSSJ2310+1855. The velocities are relative to the redshift of the absorption system in the final catalog (grey dashed line). The NIR transitions are offset by a median velocity of 17~\kms\ (red dashed line). Right: Velocity offsets after applying the 17~\kms\ shift to the NIR spectrum. There is no evidence for any systematic variation in the measured centroid velocity as a function of wavelength. The final centroid velocities vary by $\sim$~10~\kms\ which is the width of a spectral pixel.}\label{fig:dv_wavelength}
\end{figure*}

\begin{figure*}
\centering
 \includegraphics[scale=0.48, clip = True, trim = 0 10 0 0]{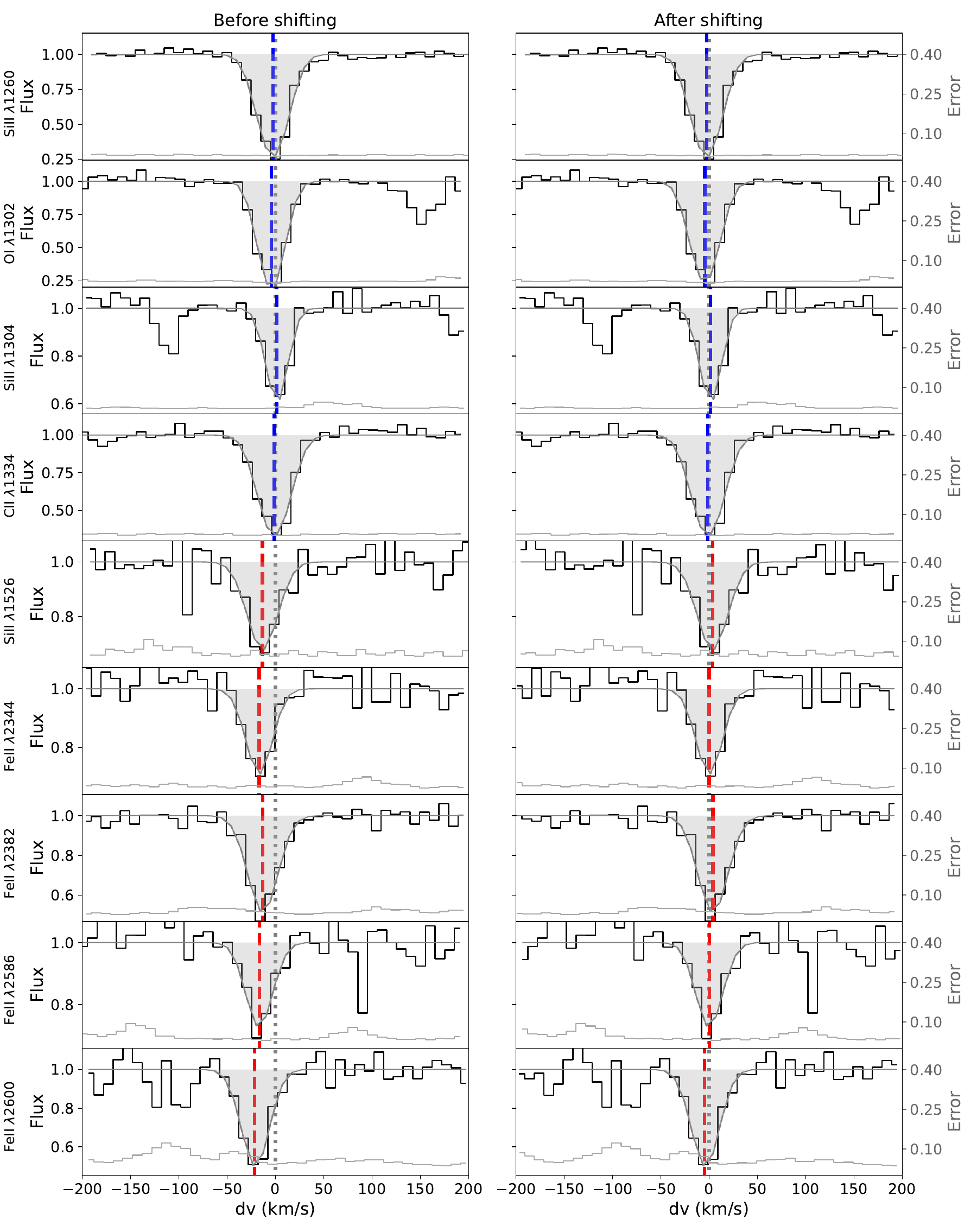}
 \caption{Profiles of VIS and NIR transitions from a single-component low-ionization absorption system at $z$ = 5.939 in the spectrum of SDSSJ2310+1855. The centroid velocities of the VIS and NIR transitions are marked with blue and red dashed lines, respectively. The velocity zero-point (grey dotted line) represents the redshift of the absorption system in the final catalog. The left-hand column shows measurements from the original spectra. The centroids of the NIR transitions are offset by $\sim$~17~\kms\ relative to the VIS transitions (see Appendix \ref{appendix:velshift} and Figure \ref{fig:dv_wavelength} for further information). The right-hand column shows the profiles after shifting the NIR spectrum by 17~\kms. The absorption profiles are very well aligned across both arms of the spectrograph, indicating that a single absolute velocity shift captures the majority of the observed centroid variation.}\label{fig:velshift_demo}
\end{figure*}

When processing the XSHOOTER spectra we found that in some cases, the redshift centroids of transitions of the same ion and absorption system falling in the VIS arm and the NIR arm (e.g. \SiII~$\lambda$1260 vs. \SiII~$\lambda$1526) were slightly offset in velocity \mbox{(5~--~20~\kms)}. This is demonstrated in the left-hand panel of Figure \ref{fig:dv_wavelength} which compares the centroid velocities of transitions falling in the VIS (blue markers) and NIR (red markers) arms for a single-component low-ionization absorption system (the line profiles of the individual transitions are shown in the left-hand column of Figure \ref{fig:velshift_demo}). The velocities are relative to the redshift of the absorption system in the final catalog (grey dashed line). There is a clear offset between the velocities of transitions falling in the VIS and NIR arms, with a magnitude of $\sim$~17\kms\ (red dashed line). 

For each quasar where such a velocity offset was identified, a single velocity shift ($\pm$~5~--~20~\kms) was applied to the whole NIR spectrum (for which the uncertainties on the wavelength solutions are slightly larger) to maximise the inter-arm alignment for all absorption systems. The right-hand panel of Figure \ref{fig:dv_wavelength} shows the relative centroid velocities of the transitions in the example system after shifting the NIR spectrum. There is no evidence for any systematic variation in the measured centroid velocity as a function of wavelength (this is also apparent from the fits to the shifted line profiles in the right-hand column of Figure \ref{fig:velshift_demo}). Higher order dispersion effects may be present but do not appear to leave any detectable signature at the spectral resolution and sampling of our observations. The final centroid velocities are spread over a $\sim$~10~\kms\ range which is the width of a spectral pixel. This is significantly larger than the error on the absolute wavelength calibration. This remaining velocity spread may instead reflect uncertainties on the line profile shape originating from factors that are not accounted for in the error budget, such as continuum normalization. 

\section{Accompanying Data Tables and Figures}

\begin{table*}
\begin{nscenter}
\begin{tabular}{l|c|c|c|c|c}
\hline
QSO ID & Sample & z & Ref & R(VIS) & R(NIR) \\ \hline
SDSSJ0927+2001  & E-XQR-30 & 5.772 & 1,2$^a$ & 12900 & 9400 \\
PSOJ065+01 & XQR-30 & 5.79 & 3 & 10700 & 9700 \\ 
PSOJ308-27 & XQR-30 & 5.7985 & 4 & 11800 & 10600 \\ 
SDSSJ0836+0054  & E-XQR-30 & 5.81  & 5  & 13100 & 10200 \\
PSOJ242-12 & XQR-30 & 5.83 & 6 & 10700 & 9700 \\ 
PSOJ025-11 & XQR-30 & 5.85 & 6 & 10500 & 9700 \\ 
PSOJ183-12 & XQR-30 & 5.86 & 6 & 11900 & 10000 \\ 
PSOJ023-02 & XQR-30 & 5.90 & 6 & 10900 & 9700 \\ 
PSOJ108+08 & XQR-30 & 5.9485 & 4 & 12200 & 9800 \\ 
PSOJ089-15 & XQR-30 & 5.957 & 4 & 12000 & 9700 \\ 
VDESJ2250-5015 & XQR-30 & 5.9767 & 4 & 10800 & 9600 \\ 
ULASJ0148+0600  & E-XQR-30 & 5.98 & 7  & 13300 & 10000 \\
PSOJ029-29 & XQR-30 & 5.984 & 6 & 10800 & 9900 \\ 
PSOJ007+04 & XQR-30 & 6.0008 & 8 & 11500 & 9700 \\ 
SDSSJ2310+1855 & XQR-30 & 6.0031 & 9 & 12700 & 9800 \\ 
PSOJ009-10 & XQR-30 & 6.0039 & 8 & 11500 & 10500 \\ 
ATLASJ029-36 & XQR-30 & 6.02 & 10 & 10100 & 9200 \\ 
SDSSJ0818+1722  & E-XQR-30& 6.02  & 11   &  11000 & 7600 \\
SDSSJ1306+0356  & E-XQR-30&  6.0337 & 8  & 12000 & 9600 \\
VDESJ0408-5632 & XQR-30 & 6.0345 & 4 & 11300 & 9700 \\ 
PSOJ158-14 & XQR-30 & 6.0679 & 12 & 10800 & 9100 \\ 
SDSSJ0842+1218 & XQR-30 & 6.0763 & 8 & 11400 & 10000 \\ 
PSOJ239-07 & XQR-30 & 6.1106 & 12 & 11300 & 11000 \\ 
CFHQSJ1509-1749 & E-XQR-30 & 6.1225 & 8 & 11800 & 8000 \\
ULASJ1319+0950   & E-XQR-30& 6.133 & 9 &  13700 & 9800 \\
PSOJ217-16 & XQR-30 & 6.1498 & 8 & 11500 & 10200 \\ 
PSOJ217-07 & XQR-30 & 6.1663 & 4 & 11200 & 9900 \\ 
PSOJ359-06 & XQR-30 & 6.1722 & 12 & 11300 & 10000 \\ 
PSOJ060+24 & XQR-30 & 6.18 & 6 & 11500 & 10300 \\ 
PSOJ065-26 & XQR-30 & 6.1877 & 8 & 11700 & 10500 \\ 
SDSSJ1030+0524  & E-XQR-30 &  6.308 & 5 & 12300 & 8400 \\
SDSSJ0100+2802  & E-XQR-30 & 6.3258 & 13   & 11400 & 10300 \\
ATLASJ025-33   & E-XQR-30 &  6.3379  & 8  & 11200 & 9300 \\ 
J2211-3206 & XQR-30 & 6.3394 & 8 & 10600 & 9100 \\ 
J1535+1943 & XQR-30 & 6.38 & 3 & 11600 & 9700 \\ 
J1212+0505 (PSOJ183+05) & XQR-30 & 6.4386 & 14 & 11900 & 10200 \\ 
J0439+1634  & E-XQR-30& 6.5188 & 15  & 9500 & 8200 \\
VDESJ0224-4711 & XQR-30 & 6.526 & 16 & 11200 & 9400 \\ 
PSOJ036+03   & E-XQR-30 &  6.5412 & 17 & 10700 & 9200 \\ 
PSOJ231-20 & XQR-30 & 6.5864 & 8 & 10800 & 9800 \\ 
PSOJ323+12 & XQR-30 & 6.5881 & 18 & 10900 & 9800 \\ 
J0923+0402 & XQR-30 & 6.6336 & 4 & 11600 & 10200 \\
\hline
\end{tabular}
\end{nscenter}
\caption{Adopted emission redshifts and spectral resolutions for the XQR-30 and archival (E-XQR-30) quasars. We adopted the best redshift measurements that were available when commencing construction of the absorption line catalog. The redshifts are consistent with those published in D'Odorico et al. (in prep) to within 300~\kms\ and these small differences do not have any significant impact on the final dataset. The spectral resolution values are estimated using seeing measurements as described in Section \ref{subsubsec:spectral_res}. References: 1) \citet{Bosman18}, 2) \citet{Wang10}, 3) \citet{Zhu21}, 4) D'Odorico et al. (in prep), 5) \citet{Kurk07}, 6) \citet{Banados16}, 7) \citet{Becker15}, 8) \citet{Decarli18}, 9) \citet{Wang13}, 10) \citet{Carnall15}, 11) \citet{Carilli10}, 12) \citet{Eilers21}, 13) \citet{Wang16}, 14) \citet{Venemans20}, 15) \citet{Yang19b}, 16) \citet{Reed19}, 17) \citet{Banados15}, 18) \citet{Mazzucchelli17}. }
$^a$This redshift is reported in \citet{Bosman18} based on the CO(2-1) measurement from \citet{Wang10}.\label{redshift_table}
\end{table*}

\begin{table*}
\begin{nscenter}
\begin{tabular}{l|c|c|c|c|c}
\hline
  & \multicolumn{5}{c}{Masked Regions (nm)} \\ 
QSO ID & \Lya\ & Telluric(Y) & Telluric (J~--~H) & Telluric (H~--~K) & Telluric(K) \\ \hline
SDSSJ0927+2001 &  & & 1345~--~1450 & 1800~--~1940 & \\ 
PSOJ065+01 & 825.0~--~827.5 & & & & \\ 
SDSSJ0836+0054 & & & 1345~--~1450 & 1800~--~1940 &  \\ 
PSOJ025-11 & 825.0~--~836.0 & & & & 1998~--~2030, 2100~--~2280  \\ 
PSOJ183-12 & 830.0~--~838.8 & & & & \\ 
PSOJ023-02 &  & 1008~--~1030 & 1345~--~1450 & 1800~--~1940 & 1998~--~2030 \\ 
PSOJ089-15 & 845~--~850 &  & & 1800~--~1940 & \\ 
J2250-5015 & 845.0~--~851.5 & & 1345~--~1485 & 1800~--~1940 & 2100~--~2280 \\ 
PSOJ007+04  & & & 1350~--~1375 & & 1998~--~2030, 2100~--~2280\\ 
SDSSJ2310+1855 & & & 1345~--~1450 & 1800~--~1940 & \\ 
ATLASJ029-36 & 853.0~--~854.5 & 1000~--~1025 & 1345~--~1450 & 1800~--~1975 & 2100~--~2280 \\ 
SDSSJ0818+1722 & & & 1345~--~1450 & 1800~--~1940 &  \\ 
SDSSJ1306+0356 & & & 1345~--~1450 & 1800~--~1940 & \\ 
J0408-5632 & & & 1350~--~1415 & 1800~--~1975 & 1998~--~2030, 2100~--~2280 \\ 
PSOJ239-07 & 865~--~880 & & & & \\ 
CFHQSJ1509-1749 & & & & 1800~--~1940 & 2100~--~2280 \\ 
PSOJ359-06 & 870~--~872.6 &  & & 1800~--~1940 & \\ 
PSOJ060+24 & & & 1345~--~1450 & & \\ 
SDSSJ1030+0524 & & & 1345~--~1450 & 1800~--~1940 & \\ 
SDSSJ0100+2802 & & & 1345~--~1450 & 1800~--~1940 & 2045~--~2080 \\ 
J2211-3206 & & & 1345~--~1450 & 1800~--~1940 & \\ 
J1535+1943 & 895.0~--~897.8 & & 1345~--~1485 & 1800~--~1975 & \\ 
J1212+0505 & & & 1350~--~1415 & 1820~--~1940 & 1998~--~2030 \\ 
J0439+1634 & & & 1350~--~1420 & 1810~--~1940 & 1998~--~2030, 2045~--~2080 \\ 
PSOJ036+03 & & & 1345~--~1450 & 1800~--~1940 & \\ 
J0923+0402 & & & 1345~--~1485 & 1800~--~1975 & \\ 
\hline
\end{tabular}
\end{nscenter}
\caption{Wavelength regions masked during system finding. The wavelength regions listed in this table were manually masked during the initial system finding due to either strong absorption just redward of the quasar \Lya\ emission line preventing the continuum level from being accurately determined, or significant contamination by skylines or telluric absorption (see discussion in Section \ref{subsec:system_search}). No spectral regions were masked for the 16 quasars not included in this table.}\label{masked_region_table}
\end{table*}

\begin{table*}
\begin{nscenter}
\setlength{\tabcolsep}{4pt} 
\begin{tabular}{c|c|c|c|c|c|c|c|c|c|c|c|c}
\hline
 &  &  &  & Rest-Frame & & Column & & & Doppler $b$ & &  & (P)roximate \\
Syst & Comp &  & Ion/ & Equivalent & & Density & & $\Delta v_{90}$ & parameter & Sample & Primary & or \\
ID & ID & Redshift & Species & Width $W$(\AA) & Flag & $\log_{10}\left(\frac{\rm N}{\rm cm^{-2}}\right)$ & Flag & (\kms) & (\kms) & Flags & Sample & (I)ntervening \\
(1) & (2) & (3) & (4) & (5-6) & (7) & (8-9) & (10) & (11) & (12-13) & (14) & (15) & (16) \\ \hline
9 & - & 4.77151 & MgII 2796 & 0.821 $\pm$ 0.013  & - & 13.53 $\pm$ 0.02  & - & 128 & - & YYYY  & Y  & I \\
- & - & 4.77151 & MgII 2803 & 0.530 $\pm$ 0.013  & - & - & - & - & - & - \\
- & 9a & 4.77078 & MgII & - & - & 13.21 $\pm$ 0.02  & - & 70 & 23.5 $\pm$  2.0 & YYYY  & Y  & I \\
- & 9b & 4.77198 & MgII & - & - & 13.22 $\pm$ 0.02  & - & 71 & 23.7 $\pm$  1.8 & YYYY  & Y  & I \\
- & 9c & 4.77399 & MgII & - & - & 12.03 $\pm$ 0.14  & - & 47 &  6.2 $\pm$  0.0 & YYYY  & Y  & I \\
9 & - & 4.77078 & FeII 2600 &  0.042  & U &  12.47  & U & 71 & - & YYNY  & N  & I \\
- & 9a & 4.77078 & FeII & - & - &  12.47  & U & 71 & 23.5 $\pm$  2.0 & YYNY  & N  & I \\
9 & - & 4.77159 & SiII 1526 & 0.082 $\pm$ 0.005  & - & 13.51 $\pm$ 0.04  & - & 112 & - & YYYY  & N  & I \\
- & 9a & 4.77078 & SiII & - & - & 13.00 $\pm$ 0.07  & - & 67 & 23.5 $\pm$  2.0 & YYYY  & N  & I \\
- & 9b & 4.77198 & SiII & - & - & 13.35 $\pm$ 0.04  & - & 67 & 23.7 $\pm$  1.8 & YYYY  & N  & I \\
9 & - & 4.77110 & AlII 1670 & 0.105 $\pm$ 0.007  & - & 12.44 $\pm$ 0.04  & - & 110 & - & YYYY  & N  & I \\
- & 9a & 4.77078 & AlII & - & - & 12.31 $\pm$ 0.04  & - & 67 & 23.5 $\pm$  2.0 & YYYY  & N  & I \\
- & 9b & 4.77198 & AlII & - & - & 11.84 $\pm$ 0.11  & - & 67 & 23.7 $\pm$  1.8 & YYYY  & N  & I \\
9 & - & 4.77197 & CIV 1548 & 0.548 $\pm$ 0.007  & - & 14.29 $\pm$ 0.02  & - & 290 & - & YYYY  & Y  & I \\
- & - & 4.77197 & CIV 1550 & 0.322 $\pm$ 0.008  & - & - & - & - & - & - \\
- & 9d & 4.76942 & CIV & - & - & 13.63 $\pm$ 0.06  & - & 102 & 40.6 $\pm$  7.4 & YYYY  & Y  & I \\
- & 9e & 4.77067 & CIV & - & - &  13.43 $\pm$ 0.09  & S & 63 & 21.6 $\pm$  5.2 & YYYY  & Y  & I \\
- & 9f & 4.77244 & CIV & - & - &  13.56 $\pm$ 0.04  & S & 89 & 34.6 $\pm$  4.3 & YYYY  & Y  & I \\
- & 9g & 4.77413 & CIV & - & - &  13.95 $\pm$ 0.02  & S & 58 & 18.8 $\pm$  1.2 & YYYY  & Y  & I \\
\hline
\end{tabular}
\end{nscenter}
\caption{A condensed excerpt of the absorber catalog for J0408-5632. Full catalogs for all quasars and a full catalog for the entire sample are available in the online supplementary material and GitHub repository. Column numbers match those in the published catalogs. Absorption components are grouped into systems as described in Section \ref{subsec:system_grouping}. For each ion in each absorption system, we provide the total $\log N$ and $\Delta v_{90}$ as well as the total $W$ of each transition. We also list the $\log N$, $\Delta v_{90}$ and $b$ parameter of each individual Voigt profile component. \mbox{(1) System} ID. (2) Component ID, consisting of a system number and a component label (letter). Components with the same ID were fit using the same $z$ and $b$ parameter (Section \ref{subsec:voigt_fitting}). (3) Optical-depth-weighted mean redshift (systems) or Voigt profile redshift (components). We do not quote errors on the redshifts which are typically on the order of 10$^{-5}$. (4) Ion or species. \mbox{(5-7) Equivalent} width measurement, error, and flag reported for each transition in each system. Measurements for ions detected at $<$~3$\sigma$ are flagged as upper limits (`U'). This quantity is not reported for individual components. (8-10) Column density measurement, error, and flag, reported for both systems and individual components. Flags indicate upper limits (for ions detected at $<$~3$\sigma$; `U'), saturated profiles (optical depth $\tau >$~2; `S'), and ions for which $\log N$ had to be fit for different transitions independently (`M'; see Section \ref{subsec:voigt_fitting}). \mbox{(11) $\Delta v_{90}$} measurement. \mbox{(12-13) $b$} parameter measurement and error, only reported for individual components. (14) Sample flags indicating whether \mbox{a) the} system was found by the automatic line finder (Section \ref{subsec:system_search}), b) the system was classified as real by at least 3/5 expert checkers (Section \ref{subsec:visual_inspection}), c) the relevant transitions lie in spectral regions free of significant sky contamination (Section \ref{subsec:system_search}), and d) the relevant transitions lie in regions free of BAL features (Section \ref{subsec:abs_path}). (15) Flag indicating whether the system is part of the primary sample (Section \ref{subsec:flags}). (16) Flag indicating whether the absorption system is proximate (P) or intervening (I) (Section \ref{subsec:proximate_intervening}).}\label{table:catalog}
\end{table*}

\begin{figure*}
\centering
 \includegraphics[scale=0.6]{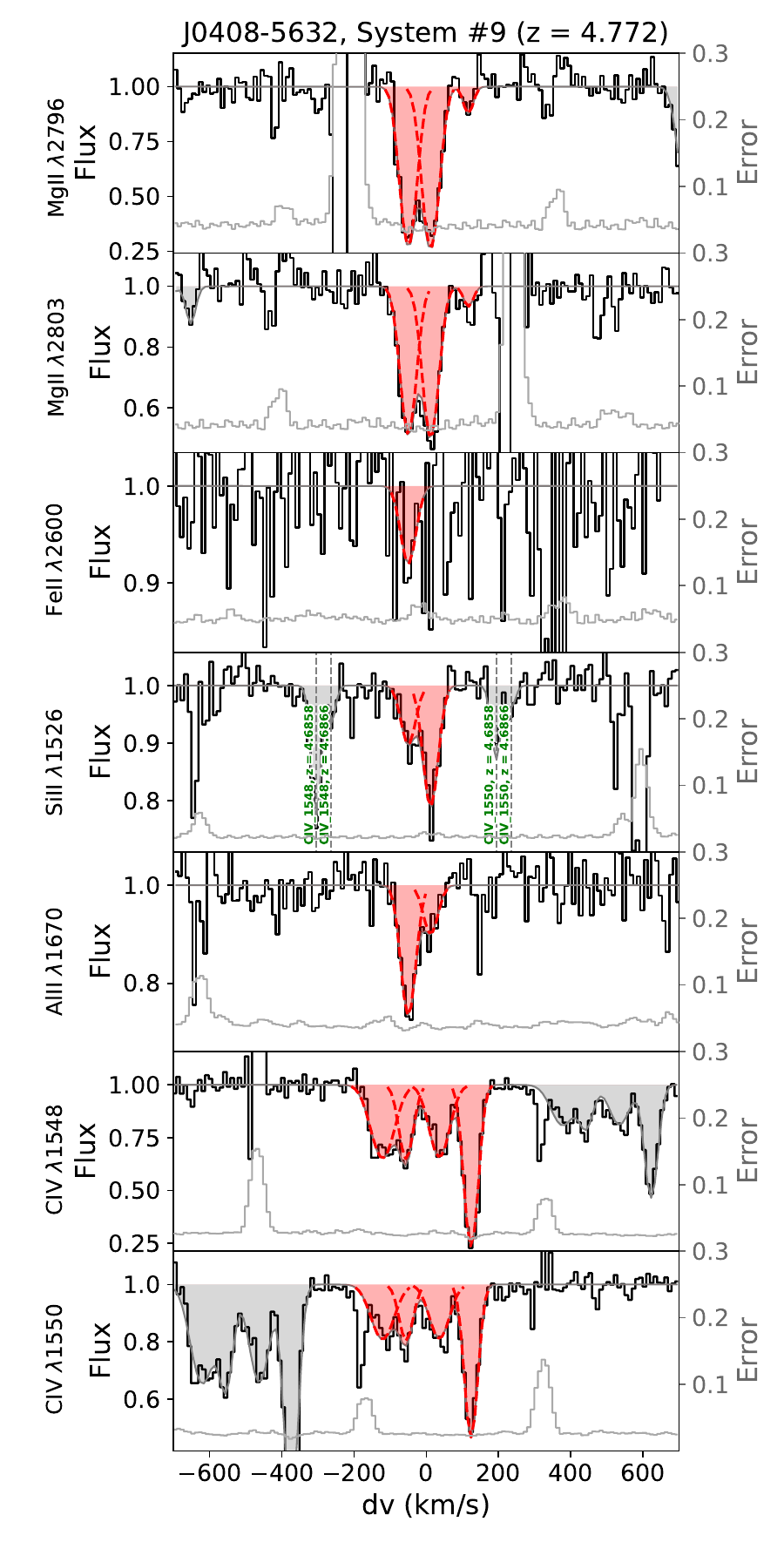}
 \caption{Voigt profile fits to the absorption system listed in Table \ref{table:catalog}. Each row shows data for a single transition. The black and grey curves show the continuum-normalized flux and error, respectively. The flux and error values are plotted over different $y$ ranges which are shown on the left and right-hand $y$-axes, respectively. The $x$-axis shows the velocity offset from the median redshift of the Voigt components. Red dashed lines indicate individual Voigt components and red filled regions show the total absorption from all components combined. Grey dashed lines and filled regions indicate absorption associated with systems at other redshifts, which are labelled with green text.}\label{fig:fitPlots}
\end{figure*}

\begin{table*}
\begin{nscenter}
\begin{tabular}{l|c|c|c}
\hline
 QSO ID & \multicolumn{3}{c}{BAL Regions (nm)} \\
  & C~\textsc{IV} & Si~\textsc{IV} & N~\textsc{V} \\ \hline
PSOJ065+01 & 918.6~--~995.0 & 865.1~--~896.7 & ~--~ \\ 
SDSSJ0836+0054 & 956.6~--~965.6 & ~--~ & ~--~ \\ 
PSOJ183-12 & 975.9~--~1010.1 & ~--~ & ~--~ \\ 
PSOJ023-02 & 981.4~--~1042.2 & ~--~ & ~--~ \\ 
PSOJ089-15 & 971.0~--~1072.2 & 883.8~--~967.5 & 848.2~--~858.5 \\ 
J2250-5015 & 900.8~--~1001.2 & 854.5~--~880.8 & ~--~ \\ 
ULASJ0148+0600 & 980.9~--~1023.0 & ~--~ & ~--~ \\ 
SDSSJ2310+1855 & 986.3~--~1016.5 & ~--~ & ~--~ \\ 
PSOJ009-10 & 937.5~--~956.7 & 846.1~--~863.2 & ~--~ \\ 
J0408-5632 & 985.1~--~1002.1 & ~--~ & ~--~ \\ 
SDSSJ0842+1218 & 995.1~--~1022.4 & ~--~ & ~--~ \\ 
PSOJ239-07 & 1079.9~--~1099.4 & 983.6~--~992.0 & 865.9~--~880.3 \\ 
CFHQSJ1509-1749 & 997.2~--~1017.7 & ~--~ & ~--~ \\ 
PSOJ217-07 & 947.7~--~1029.0 & 885.1~--~900.3 & ~--~ \\ 
J2211-3206 & 1043.7~--~1100.8 & 959.6~--~992.3 & ~--~ \\ 
J0439+1634 & 1079.4~--~1120.5 & 991.2~--~1010.0 & ~--~ \\ 
PSOJ231-20 & 1162.2~--~1171.5 & ~--~ & 931.2~--~938.6 \\ 
J0923+0402 & 1065.3~--~1157.7 & 979.8~--~1044.6 & ~--~ \\ 
\hline
\end{tabular}
\end{nscenter}
\caption{Wavelength regions showing broad absorption line (BAL) features. BAL regions were identified by searching for contiguous troughs with normalized flux less than 0.9 over a velocity interval of at least 2000~\kms\ \citep{Bischetti22}. The \CIV\ and \SiIV+\NV\ BAL regions are computed directly from the velocity intervals and redshifts published in \citet{Bischetti22} and Bischetti et al. (submitted), respectively.}\label{bal_region_table}
\end{table*}

\label{lastpage}

\end{document}